\documentclass[reprint,amsmath,amssymb,aps,prx]{revtex4-2}

\usepackage{graphicx,tikz,tikz-3dplot}
\usepackage[dvipsnames]{xcolor}
\usetikzlibrary{math,patterns,decorations.pathreplacing}
\graphicspath{\FIGS}
\usepackage{dcolumn}
\usepackage{bm}
\usepackage{hyperref}
\usepackage{braket}
\usepackage{booktabs} 
\usepackage{siunitx}
\usepackage{soul}

\newcommand{\etc}{\textit{etc}.}

\newcommand{\kp}{\bm{k}}
\newcommand{\tk}{\theta_{\kp}}
\newcommand{\Ek}{\varepsilon_{\bm{k}}}
\newcommand{\Vk}[1]{V_{\bm{k}}^{\sigma_{#1}}}
\newcommand{\Wk}[1]{W^{\chi_{#1}}}
\newcommand{\Dk}{\Delta_{\bm{k}}}
\newcommand{\di}{\mathrm{d}}

\newcommand{\Tr}{ \mathrm{Tr}\, }

\newcommand{\im}{\,\mathrm{Im}\, }

\newcommand{\expval}[1]{\langle #1\rangle}

\begin{document}

\title{Transport in close-packed solids with stacking defects}

\author{C.~M.~Wilson}
\email{cw14mi@brocku.ca}
\author{R.~Ganesh}
\email{r.ganesh@brocku.ca}
\author{K.~V.~Samokhin}
\email{ksamokhin@brocku.ca}
\affiliation{Department of Physics, Brock University, St.~Catharines ON, L2S 3A1 Canada}
\date{\today}

\begin{abstract}

Lithium and sodium are the only solids that are known to lose crystalline order upon cooling. The seemingly-disordered low-temperature phase shows signatures of various close-packed structures.
The lack of order has been attributed to a hidden gauge symmetry that arises when electrons from one layer can hop to a neighbouring layer but not further. It makes all close-packed structures nearly degenerate and leads to ``structural frustration''. 
In this article, we examine whether this symmetry is reflected in transport signatures. Taking advantage of in-plane translational periodicity, we map the bulk Bloch Hamiltonian to an effective one-dimensional chain, with stacking disorder mapping to random phases of the hopping amplitudes. We derive an explicit analytic form for the Green's function of electrons and use it to calculate conductance of a bulk crystal. When hopping in the effective one-dimensional chain is restricted to nearest neighbours, conductance is completely insensitive to phase disorder, which indicates that all close-packed structures exhibit the same conductance. 
We show that the leading correction that can differentiate between close-packed structures arises from hopping to the next-nearest-neighbour layer, equivalent to second-neighbour hopping in the chain model. This process appears when a pair of next-neighbour layers are aligned in a certain way, e.g., at an hcp-like stacking fault within an fcc background. 
With this hopping included, conductance becomes sensitive to the precise arrangement of layers. When multiple stacking faults are present, the conductance decreases with increasing system size, as expected from Anderson localization.   
Our results are applicable to pressurized lithium and sodium, where conductance measurements can identify and characterize stacking faults. 

\end{abstract}

\maketitle

\section{\label{sec:introduction}Introduction}

A general rule in physical systems is that decreasing temperature leads to increased order. In the context of solids, this is prominently seen in liquid-to-crystal transitions upon cooling. Lithium and sodium are the only two known exceptions to this rule. Both metals form highly ordered bcc structures at room temperature and ambient pressure. When cooled, they undergo a martensitic transition -- at 77 K in the case of lithium~\cite{barrett_Li_1947} and at 36 K in the case of sodium~\cite{Barrett1955,Barrett1956}. Below these temperatures, they show simultaneous signatures of multiple structures, in addition to amorphous-like diffuse peaks. Crucially, the signatures seen correspond to various close-packed structures: fcc~\cite{Ackland2017}, hcp~\cite{McCarthy1980}, 9R~\cite{Overhauser1984,Schwarz1991}, dhcp~\cite{Witt2021}, etc. In the case of lithium, it has been suggested that the true ground state corresponds to an fcc structure, but it can only be accessed via a specific protocol~\cite{Ackland2017}. Two of the present authors have proposed an explanation for the disordered nature of the low-temperature phase~\cite{he2025}. The key idea is that lithium and sodium are close to an idealized point where a hidden gauge symmetry makes all close-packed structures degenerate. This leads to structural frustration where the solid is unable to ``choose'' one structure out of an infinite nearly-degenerate set. In this article, we consider the consequences of this near-degeneracy for electronic transport. 

The concept of ``accidental'' degeneracy is well known in the field of frustrated magnetism~\cite{MoessnerRamirez2006}. Prominent examples include spin ice~\cite{Gingras2011SpinIce} and spin glasses~\cite{Binder1986}. It refers to a system with a large ground-state degeneracy that does not arise from a true symmetry. For example, the material MnSc$_2$S$_4$ hosts a large family of spiral ground states~\cite{Gao2017}, which are not related to one another by any symmetry of the Hamiltonian. Rather, they arise due to frustration where all terms in the Hamiltonian cannot be minimized independently. The ground states are compromise configurations that carry the lowest energy in this setting. Without protection from a symmetry, their degeneracy is very fragile. Small effects such as spin-wave zero-point energies, thermal fluctuations, or further-neighbour interactions can break the degeneracy and select an ordered state. 
Careful experiments may be required to identify the ordered state, or indeed to determine if ordering has taken place, e.g., see Refs.~\cite{Krimmel2006,Kalvius2006,Mücksch_2007,Iqbal2018} in the context of MnSc$_2$S$_4$. 

Here, we examine a related idea of the structural frustration in a close-packed solid. We consider a frustrated regime where an infinite family of structures are nearly degenerate. We identify the leading correction that can break the degeneracy. 
Finally, we examine if transport measurements can detect structural (dis)ordering and if so, be used to characterize the structure of a sample. 

Transport in disordered solids is known to be affected by Anderson localization~\cite{Lee_TVR_1985}. In low dimensions, an infinitesimal amount of disorder suffices to localize electrons and to render the system insulating. This is due to destructive interference from repeated scattering processes. This notion has been studied in various settings (e.g., on various lattices) and with various forms of disorder (e.g., with a random potential drawn from a uniform distribution). Disorder may also appear in hopping amplitudes, local coordination number, etc. A particularly exciting form is phase disorder, where the electron picks up a random phase as it hops from one site to the next~\cite{Altland1999}. Here, we study phase disorder in a one-dimensional (1D) system -- as an effective model for close-packed solids. We show that, with nearest-neighbour hoppings, the physical properties, including transport, are completely insensitive to the disorder. We identify the leading correction that makes the disorder relevant and leads to localization of electron wave functions in the stacking direction. 

A standard approach to calculating transport in low-dimensional systems is the Landauer formalism~\cite{nazarov_blanter_2009}.
A key idea in this work is that this same formalism can be used to calculate the transport properties of a stacked three-dimensional (3D) solid. We discuss close-packed structures that are constructed by stacking triangular layers, with independent lateral shifts. The resulting structure has translational symmetry in the plane perpendicular to the stacking axis, allowing for the in-plane momentum to serve as a good quantum number. Each value of the in-plane momentum labels an independent~1D channel for electronic transport along the stacking direction. 

The remainder of this article is organized as follows. In Sec.~\ref{sec:disordered_chain}, we discuss a toy problem of a 1D chain with phase disorder in the hopping coefficients. Section~\ref{sec:close_packed_solids} presents an introduction to close-packing, followed by a tight-binding model for electrons in a close-packed solid. It also demonstrates how the~3D Bloch Hamiltonian can be mapped onto the 1D chain model. In Sec.~\ref{sec:stacking_fault_transmission}, we discuss a simple stacking fault and its transport signatures. In Sec.~\ref{sec:random_chain}, we discuss a close-packed solid with random stacking, mapping it to a phase-disordered 1D chain. Section~\ref{sec:Landauer} incorporates the results of the previous sections into the Landauer formalism to calculate the resistance of a bulk sample, where every in-plane momentum corresponds to a transport channel. We conclude with a summary and discussion in Sec.~\ref{sec:conclusions}. Throughout this paper, we denote the electron charge by $-e$.

\section{\label{sec:disordered_chain}Phase-disordered chain}

In this section, we introduce the tight-binding model of a 1D phase-disordered chain. We present closed-form expressions for its eigenvalues and eigenvectors, which we use to calculate the electron Green's function in the thermodynamic limit. Later on, we will show that the results derived here can be applied to a close-packed 3D solid with an arbitrary stacking sequence.

\subsection{\label{ssec:chain_tight_binding_model}Tight-binding model}
We consider a chain containing $N$ atomic sites with open boundary conditions. We retain only nearest-neighbour hopping with amplitude $t$, and denote the on-site energy by $\varepsilon_0$ (we do not set $\varepsilon_0=0$ for reasons explained in Sec.~\ref{ssec:close_packed_mapping}). Each bond is associated with an arbitrary phase, with the electron wave function accruing a phase of $\phi_{m}$ when hopping from atom $m$ to $m+1$. Such phases may arise, for example, by placing the chain in a spatially-inhomogeneous magnetic field~\cite{peierls1933}. The setup is depicted in Fig.~\ref{fig:phase_disordered_chain}. 

The tight-binding Hamiltonian is given by a tridiagonal matrix
\begin{figure}[b]
    \centering
    \begin{tikzpicture}
    \def\n{3}
    \node at (-1,0) {$\cdots$};
    \fill (-2,0) circle (0.1) node [below,shift={(0,-0.1)}] {$1$};
    \node at (4,0) {$\cdots$};
    \fill (0,0) circle (0.1) node [below,shift={(0,-0.1)}] {$m$};
    \fill (1.5,0) circle (0.1) node [below,shift={(0,-0.1)}] {$m+1$};
    \fill (3,0) circle (0.1) node [below,shift={(0,-0.1)}] {$m+2$};
    \fill (5,0) circle (0.1) node [below,shift={(0,-0.1)}] {$N$};
    \draw [->, >=stealth, thin, gray, shift={(0,0.2)}] (0.1,0) to [out=45,in=135] (1.4,0);
    \draw [->, >=stealth, thin, gray, shift={(0,0.2)}] (1.6,0) to [out=45,in=135] (2.9,0);
    \node [black] at (0.75,0.75) { $-te^{i\phi_{m}}$};
    \node [black] at (2.25,0.75) {\small $-te^{i\phi_{m+1}}$};
\end{tikzpicture}
    \caption{1D tight-binding chain with open boundary conditions. Hopping amplitudes (with phases) between neighbouring atoms are indicated above the arrows.}
    \label{fig:phase_disordered_chain}
\end{figure}
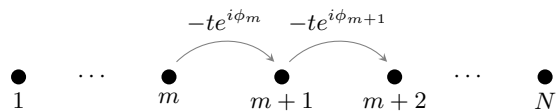
\begin{equation}
\hat{H}=\begin{pmatrix}
\varepsilon_0 & -t e^{i\phi_{1}} &  &  &  \\
- t e^{-i\phi_{1}} & \varepsilon_0 & \ddots &  &  \\
 & \ddots & \ddots & \ddots &  \\
 &  & \ddots & \varepsilon_0 & -t e^{i\phi_{N-1}} \\
 &  &  & -t e^{-i\phi_{N-1}} & \varepsilon_0
\end{pmatrix}.
    \label{eq:chain_hamiltonian}
\end{equation}
In Appendix~\ref{app:gauge_transformation},
we show that the $\phi_m$'s in this Hamiltonian can be removed by a unitary transformation, leaving the rest of the matrix untouched. That is, the Hamiltonian becomes a tridiagonal Toeplitz matrix, with the well-known eigenvalues
\begin{equation}
    \epsilon_j=\varepsilon_0+2t\cos{\beta_j},
    \label{eq:chain_eigenvalues}
\end{equation}
and eigenvectors
\begin{equation}
    \psi_j(m)=\sqrt{\frac{2}{N+1}}\,\sin(m\beta_j),
    \label{eq:chain_eigenvectors}
\end{equation}
where $\psi_j(m)$ denotes the $m^\mathrm{th}$ component of the $j^\mathrm{th}$ eigenvector, and $\beta_j$ is given by 
\begin{equation}
    \beta_j=\frac{j\pi}{N+1},
    \label{eq:chain_wavevector}
\end{equation}
with $1\leq j\leq N$ and $1\leq m\leq N$ (see, e.g., Ref.~\cite{kouachi2008}).
Using the Lagrange trigonometric identities~\cite{whittaker_watson_2021}, we may readily verify that the eigenvectors in  Eq.~(\ref{eq:chain_eigenvectors}) comprise an orthonormal set. For the remainder of this section, we work in the basis where the phases $\phi_m$'s have been removed from the Hamiltonian.

\subsection{\label{ssec:chain_greens_function}Green's function}

We introduce the Green's function, $\hat g(z)=(z-\hat{H})^{-1}$, where $z$ is a complex parameter with the dimension of energy and $\hat{H}$ is given by Eq.~(\ref{eq:chain_hamiltonian}). It is possible to calculate $\hat g(z)$ by expanding it in a series of the eigenvectors (\ref{eq:chain_eigenvectors}), according to
\begin{equation}
    g_{mn}(z)
    =\sum_{j=1}^N\frac{\psi_j(m)\psi_j^*(n)}{z-\epsilon_j}
    \label{eq:chain_eigenfunction_expansion},
\end{equation}
see, e.g., Ref.~\cite{economou2005}.
Before going further with Eq.~(\ref{eq:chain_eigenfunction_expansion}), we pause to note a useful symmetry of the Green's function: by substituting Eqs.~(\ref{eq:chain_eigenvalues}) and~(\ref{eq:chain_eigenvectors}) into Eq.~(\ref{eq:chain_eigenfunction_expansion}), it follows that
\begin{equation}
    g_{N+1-m,\,N+1-n}(z)=g_{mn}(z).
    \label{eq:chain_greens_function_symmetry}
\end{equation}
We will exploit this property in Sec.~\ref{sec:stacking_fault_transmission}, in order to extract the matrix element $g_{11}$ to calculate transport properties.

In order to evaluate the right-hand side of Eq.~(\ref{eq:chain_eigenfunction_expansion}), we work in the thermodynamic limit $N\gg1$, wherein the discrete sum over modes $j$ becomes an integral over a quasi-continuous variable, $\beta_j\rightarrow \beta\in(0,\pi)$. Since we are primarily interested in the case where the phase-disordered chain is used in transport calculations as a semi-infinite lead, this is the regime of interest to us. However, it is worthwhile to note that the inverse of a tridiagonal Toeplitz matrix, e.g., Eq.~(\ref{eq:chain_hamiltonian}) without the $\phi_{m}$'s, has a closed-form solution for arbitrary $N$~\cite{meurant_SIAM_1992}.

Assuming, with no loss of generality, that $m\geq n$, Eq.~(\ref{eq:chain_eigenfunction_expansion}) becomes, as $N\gg1$,
\begin{equation}
\label{eq:g-mn-z}
    g_{mn}(z)
    =\frac{2}{\pi}\int_0^\pi d\beta\, \frac{\sin(m\beta)\,\sin(n\beta)}{z-\epsilon(\beta)},
\end{equation}
where $\epsilon(\beta)=\varepsilon_0+2t\cos{\beta}$, with $\beta\in(0,\pi)$, is the continuum generalization of Eq.~(\ref{eq:chain_eigenvalues}). The integral in Eq. (\ref{eq:g-mn-z}) is evaluated in Appendix~\ref{app:GF_derivation} for arbitrary values of the parameter $z$. 

The retarded Green's function, which plays an essential role in transport calculations, is obtained by putting $z=E+i0^+$, where $E$ is real-valued and satisfies $\varepsilon_0-2t\leq E\leq \varepsilon_0+2t$. From Appendix~\ref{app:GF_derivation}, we obtain
\begin{equation}
    g^{\mathrm{r}}_{mn}(E)
    =\frac{1}{t}\,\frac{e^{-im\theta}\sin{\left(n\theta\right)}}{\sin{\theta}},
    \label{eq:chain_retarded_greens_function}
\end{equation}
where the superscript~`$\mathrm{r}$' denotes the retarded Green's function, and $\theta\in(0,\pi)$ is an implicit function of $E$ through 
\begin{equation}
\label{eq:chain-energy-parametrization}
    E(\theta)=\varepsilon_0+2t\cos{\theta}.
\end{equation}
Henceforth, we will deal only with the retarded Green's function and omit the superscript (with the exceptions of Appendices~\ref{app:GF_derivation} and~\ref{app:Caroli}). To return to the original basis containing the $\phi_{m}$'s, we need only to undo the unitary transformation introduced in Appendix~\ref{app:gauge_transformation}, which yields a prefactor in the Green's function:
\begin{equation}
    g_{mn}(E)
    =\frac{1}{t}\left[\prod_{r=n}^{m-1}e^{-i\phi_{r}}\right]\frac{e^{-im\theta}\sin{\left(n\theta\right)}}{\sin{\theta}}.
    \label{eq:chain_retarded_greens_function_with_phases}
\end{equation}

It is apparent from Eq.~(\ref{eq:chain_retarded_greens_function_with_phases}) that any observable which is a function only of the diagonal matrix elements of $\hat{g}(E)$ is completely insensitive to the phase disorder. For example, the density of states (DoS) per site is
\begin{align}
\nonumber    D(E)
    &=-\frac{1}{N\pi} \im \left[\mbox{Tr}\,\hat g(E)\right]=\frac{1}{N\pi t}\sum_{m=1}^N\frac{\sin^2(m\theta)}{\sin{\theta}}\\
    &=\frac{1}{4N\pi t}\,\frac{\left(2N+1\right)\sin{\theta}-\sin{\left[\left(2N+1\right)\theta\right]}}{\sin^2{\theta}}
\end{align}
(the final line above follows from Ref.~\cite[Eq.~(1.351.1)]{gradshteyn_ryzhik_2014}). We note that, since we have already assumed $N\gg1$, the second term in the numerator of $D(E)$ is negligible compared to the first, whence the DoS per site can be approximated by
\begin{equation*}
    D(E)
    \sim\frac{1}{\pi}\,\frac{1}{\sqrt{4t^2-(E-\varepsilon_0)^2}},
\end{equation*}
which is the familiar result for a disorder-free chain with periodic boundary conditions~\cite{economou2005}. As expected in the $N\rightarrow\infty$ limit, the DoS is the same for open and periodic boundary conditions. In Sec.~\ref{sec:stacking_fault_transmission} below, we consider another observable, the transmission function, that depends only on the diagonal entries of $\hat g(E)$ .

\section{\label{sec:close_packed_solids}Close-packed solids}

\subsection{\label{ssec:close_packed_geometry}Representations and geometry}

The geometry of close-packed solids and their representations have been extensively discussed elsewhere, see, for instance, Refs.~\cite{KrishnaPandey1981} and \cite{he2025}. We briefly recapitulate the details in this section.

\subsubsection{\label{sssec:close_packed_representations}Representations of close-packed solids}

Any close-packed solid can be assembled by stacking layers of triangular lattices along a common axis. Each layer must be shifted laterally relative to the layer below, in order to maintain maximal density, corresponding to a packing fraction of $\pi/3\sqrt{2}\approx 0.74$. This results in three distinct lateral positions, conventionally denoted as $A$, $B$, and $C$. Alternatively, one may introduce ``chirality'' variables $\sigma_m$, which take values $\sigma_m\in\{+,-\}$. The value `$+$' denotes ``forward'' inter-layer shifts ($A\rightarrow B,B\rightarrow C,C\rightarrow A$), while `$-$' denotes ``backward'' shifts ($A\rightarrow C,B\rightarrow A,C\rightarrow B$). Any close-packed solid can then be expressed as either a string of letters $A,B,C$ (where adjacent letters must be different), or a string of signs. The former is called a Barlow sequence, while the latter is called a H\"{a}gg code~\cite{Hagg1943}. For example, the fcc lattice with periodic boundary conditions can be represented either as $(ABC)$, a repeating sequence of three letters, or as $(+)$, a string where all chirality variables are equal to~`$+$'.

\subsubsection{\label{sssec:close_packed_local_environment}Local geometry of close-packed solids}

Any $N$-layer close-packed solid (with open boundaries along the stacking direction) is conveniently described using a hexagonal unit cell and the primitive lattice vectors
\begin{equation}
    \bm{a}=\hat{\bm{x}},\quad
    \bm{b}=-\frac{\hat{\bm{x}}}{2}+\frac{\sqrt{3}\,\hat{\bm{y}}}{2}.
    \label{eq:lattice_vectors}
\end{equation}
Here, $\bm{a}$ and $\bm{b}$ are the intra-layer, or ``in-plane'' vectors. They are both perpendicular to $\bm{c}=\hat{\bm{z}}$, the stacking direction. All lengths are measured in units of the interatomic spacing. In a solid with $N$ layers, the hexagonal unit cell contains $N$ atoms, one from each layer, at locations to be described below.

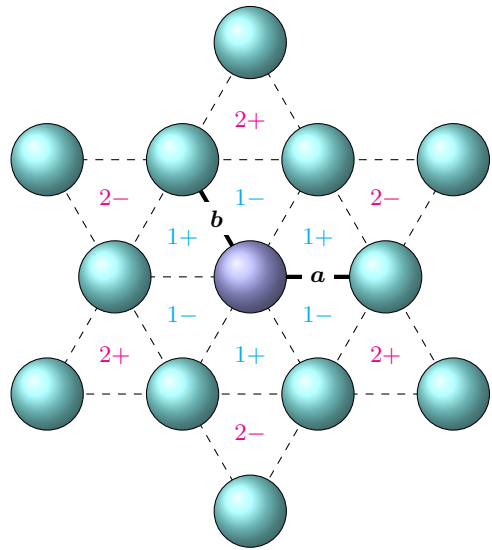
\begin{figure}[t]
    \centering
    \begin{tikzpicture}[scale=0.8]
    \def\a{2.25}
    \def\r{0.6}
    \foreach \i in {0,60,...,300} {
        \draw [dashed] (\i:\a) --++ ({\i+60}:\a) -- ({\i+60}:\a);
        \draw [dashed] (0,0) -- (\i:\a) --++ ({\i+120}:\a);
    }
    \draw [ultra thick,>=stealth,->] (0,0) -- (0:\a) node [midway,fill=white] {$\bm{a}$};
    \draw [ultra thick,>=stealth,->] (0,0) -- (120:\a) node [midway,fill=white] {$\bm{b}$};
    \foreach \i in {0,60,...,300} {
      \shadedraw[ball color=cyan!40,opacity=0.6] (\i:\a) circle (\r);
      \shadedraw[ball color=cyan!40,opacity=0.6] ({\i+30}:{sqrt(3)*\a}) circle (\r);
    }
    \shadedraw[ball color=blue!30,opacity=0.8] (0,0) circle (\r);
    \foreach \i in {30,150,270} {
      \node [cyan] at (\i:{\a/sqrt(3)}) {$1+$};
      \node [magenta] at (\i:{2*\a/sqrt(3)}) {$2-$};
      \node [cyan] at ({\i+60}:{\a/sqrt(3)}) {$1-$};
      \node [magenta] at ({\i+60}:{2*\a/sqrt(3)}) {$2+$};
    }
\end{tikzpicture}
    \caption{Atomic environment of a representative atom (center; shown in blue) within one layer of any close-packed structure. The~1nn atoms in the layer above are located at the sites marked~`$1+$' if the corresponding chirality variable in the H\"{a}gg code is~`$+$', or at the sites marked~`$1-$' if the corresponding chirality variable in the H\"{a}gg code is~`$-$'.
    In the same way,~2nn sites are~`$2+$' or~`$2-$', depending on the H\"{a}gg code entry.}
    \label{fig:in_plane_atomic_environment}
\end{figure}

A representative atom~``$0$'' in an arbitrary close-packed stacking has precisely twelve nearest neighbours (1nn), with their set denoted as $\mathcal{N}_1$. Six of the~1nn atoms are located in the same plane as~``0'', situated at the vertices of a regular hexagon with~``0'' at its centre; we denote this subset of $\mathcal{N}_1$ by $\mathcal{N}_1^{\parallel}$ (see Fig.~\ref{fig:in_plane_atomic_environment} and Table~\ref{tab:nearest_neighbours}). Of the remaining~1nn atoms, three sit in the layer immediately above and three in the layer below. Their in-plane coordinates correspond to centroids of the triangles formed by~``0'' and the $\mathcal{N}_1^{\parallel}$. In Fig.~\ref{fig:in_plane_atomic_environment}, these triangles either point up (those marked by~`$1+$'), or down (marked by~`$1-$'). If the corresponding chirality variable in the H\"{a}gg code is $\sigma_m=+$, the~1nn atoms find themselves at the sites~`$1+$', and if $\sigma_m=-$, at the sites~`$1-$'. 

Beyond the first neighbours, each atom has precisely six second-nearest neighbours (2nn). Three lie in the adjacent layer above that containing ``0'', while three lie in the layer below.
Their in-plane coordinates are given by the sites~`$2+$' or~`$2-$' in Fig.~\ref{fig:in_plane_atomic_environment}, depending on the H\"{a}gg code entry.

First and second neighbours are qualitatively the same across all close-packed structures --- they appear at the same distances, with the same number of neighbours and similar displacement vectors. In contrast, third neighbours are qualitatively different across structures. 
To describe the third-nearest neighbours (3nn), we introduce new variables $\chi_m$, which describe a pair of contiguous chirality variables,
\begin{equation}
    \chi_m=\left(\sigma_m\sigma_{m+1}\right).
    \label{eq:close_packed_chi_definition}
\end{equation}
Given a H\"{a}gg code of length $N$, one can form $N-1$ pairs of the $\chi$ variables according to the prescription in Eq.~(\ref{eq:close_packed_chi_definition}). The rule for identifying the~3nn atoms can be formulated as follows: if $\chi_m\in\{(+-),(-+)\}$, then atom~``0'' of layer $m$ has a 3nn in layer $m+2$. This third neighbour is located at a position $\sqrt{8/3}\,\bm{\hat{z}}\approx 1.63\bm{\hat{z}}$ relative to~``0'' (i.e., along the stacking axis). If $\chi$ is neither $(+-)$ nor $(-+)$, then there is no atom at this position. The next closest neighbour is farther away, at a distance of $\sqrt{3}\approx 1.73$. In this work, we shall say that an atom has a~3nn only if the former condition is met. That is, we consider hoppings up to a distance of $\sqrt{8/3}$ (in units where the interatomic spacing is set to unity) and not farther.

\begin{table}[t]
    \centering
    \begin{tabular}{c c}
        \toprule
        Neighbours&\quad Representative coordinate\\[0.3em]\midrule
        $\mathcal{N}_1^{\parallel}$ 
        & $(1,0,0)$  \\[0.3em]
        $\mathcal{N}_{1}^{\perp,+}$
        & $(\frac12,\frac{1}{2\sqrt{3}},\sqrt{\frac23})$ \\[0.3em]
        $\mathcal{N}_{2}^{\perp,+}$
        & $(0,\frac{2}{\sqrt{3}},\sqrt{\frac23})$ \\[0.3em]
        $\mathcal{N}_{3}$
        & $(0,0,2\sqrt{\frac23})$\\[0.3em]\bottomrule
    \end{tabular}
    \caption{Sets of the nearest-neighbour vectors. The order of a neighbour (first, second, \etc) is indicated by the subscript. The sets $\mathcal{N}_{1}^{\perp,-}$ and $\mathcal{N}_{2}^{\perp,-}$ may be obtained from $\mathcal{N}_{1}^{\perp,+}$ and $\mathcal{N}_{2}^{\perp,+}$ respectively by inverting the in-plane coordinates.}
    \label{tab:nearest_neighbours}
\end{table}

\subsection{\label{ssec:close_packed_tight_binding}3D tight-binding model}

We consider an electron in an $N$-layer close-packed solid with an arbitrary stacking sequence. All hopping processes are retained up to the third-nearest-neighbor (wherever~3nn atoms are actually present). Periodic boundary conditions are enforced along $\bm{a}$ and $\bm{b}$, therefore the two-dimensional in-plane momentum $\kp$ is a good quantum number. We use open boundary conditions along the stacking direction, $\bm{c}$. 

In the mixed basis $(\kp,m)$, where $\kp$ denotes in-plane momentum and $m$ indexes the layers, the Hamiltonian is represented by an $N\times N$ matrix. Its diagonal entries, denoted as $\Ek$, encode hopping processes within a single layer, i.e., from~``0'' to any of the $\mathcal{N}_1^{\parallel}$ atoms. We have
\begin{equation}
    \Ek=-t_1\sum_{\bm{\eta}\,\in\,\mathcal{N}_1^{\parallel}}e^{-i\kp\cdot\bm{\eta}},
    \label{eq:Ek}
\end{equation}
where $t_1$ is the~1nn hopping parameter. Inter-layer hopping processes between~``0'' and its~1nn and~2nn in the adjacent layers are accounted for by off-diagonal matrix elements $\Vk{m}$, given by
\begin{equation}
    \Vk{m}
    =-t_1\sum_{\bm{\eta}\,\in\,\mathcal{N}_{1}^{\perp,\pm}}e^{-i\kp\cdot\bm{\eta}}
    -t_2\sum_{\bm{\eta}\,\in\,\mathcal{N}_{2}^{\perp,\pm}}e^{-i\kp\cdot\bm{\eta}},
    \label{eq:Vk}
\end{equation}
where $t_2$ is the~2nn hopping parameter. The summation runs over $(\mathcal{N}_1^{\perp,+},~\mathcal{N}_2^{\perp,+})$ or $(\mathcal{N}_1^{\perp,-},~\mathcal{N}_2^{\perp,-})$, depending on whether $\sigma_m=+$ or $-$. Lastly, the 3nn couplings lead to the following entries in the second off-diagonal of the Hamiltonian:
\begin{equation}
    \Wk{m}=\left\{ 
    \begin{array}{ll}
    -t_3, & \mathrm{if}~\chi_m\in \{(+-),(-+)\},\\
    0,& \mathrm{otherwise}.
    \end{array}\right.
    \label{eq:Wk}
\end{equation}
Here, $t_3$ is the~3nn hopping parameter. Note that $\Wk{m}$ is independent of $\kp$, due to the fact that the displacement vectors from~``0'' to the~3nn atoms are along $\bm{c}$, i.e., perpendicular to the in-plane momentum $\kp$. We proceed taking $t_1$, $t_2$, and $t_3$ to be real-valued. In lithium and sodium, this is justified as the $t$'s arise from overlaps of $s$ orbitals. 

Combining Eqs.~(\ref{eq:Ek}) through~(\ref{eq:Wk}), we arrive at the following Bloch Hamiltonian:
\begin{equation}
    \hat{H}_{\bm{k}}=\begin{pmatrix}
    \Ek & \Vk{1} & \Wk{1} &  &   &   \\[0.3em]
    {\Vk{1}}^* & \Ek & \ddots & \ddots &   &   \\[0.3em]
    \Wk{1} & \ddots & \ddots & \ddots & \ddots &   \\[0.3em]
    & \ddots & \ddots & \ddots & \Vk{N-2} & \Wk{N-2} \\[0.3em]
    &  & \ddots & {\Vk{N-2}}^* & \Ek & \Vk{N-1} \\[0.3em]
    &  &  & \Wk{N-2} & {\Vk{N-1}}^* & \Ek
\end{pmatrix}.
\label{eq:close_packed_hamiltonian}
\end{equation}
The off-diagonal $V$ terms have the following crucial symmetry:
\begin{equation}
    V_{-\kp}^{\sigma_m}
    =V_{\kp}^{-\sigma_m}.
    \label{eq:close_packed_Vk_symmetry}
\end{equation}
To see this, we replace $\kp\to-\kp$ in Eq.~(\ref{eq:Vk}). Then each exponential is taken into 
\begin{equation*}
    e^{-i\kp\cdot\bm{\eta}}
    \to
    e^{-i\left(-\kp\right)\cdot\bm{\eta}}
    =e^{-i\kp\cdot\left(-\bm{\eta}\right)}.
\end{equation*}
But $\bm{\eta}\to-\bm{\eta}$ takes the elements of $\mathcal{N}_{1}^{\perp,+}$ (or $\mathcal{N}_{2}^{\perp,+}$) into $\mathcal{N}_{1}^{\perp,-}$ (or $\mathcal{N}_{2}^{\perp,-}$), since one set can be obtained from the other by inverting the in-plane coordinates. This can be seen by comparing the positions of sites~`$1\pm$' and~`$2\pm$' in Fig.~\ref{fig:in_plane_atomic_environment}.
As a corollary of Eq.~(\ref{eq:close_packed_Vk_symmetry}),
\begin{equation}
    \left(V_{\kp}^{-\sigma_m}\right)^*
    =V_{-\kp}^{-\sigma_m}
    =\Vk{m},
    \label{eq:close_packed_Vk_complex_conjugate}
\end{equation}
whence $V_{\kp}^{-\sigma_m}$ and $\Vk{m}$ are complex conjugates. 

\begin{figure*}[htbp]
    \centering
    \definecolor{col1}{HTML}{dabfff} 
\definecolor{col2}{HTML}{907ad6} 
\definecolor{col3}{HTML}{4f518c} 
\definecolor{col4}{HTML}{2c2a4a}
\tdplotsetmaincoords{52}{135}     
\begin{tikzpicture}[tdplot_main_coords,scale=1.0]
  \def\d{3}
  \def\f{1.32}


  \tikzset{pics/A/.style n args={3}{code={
  \foreach \t in {0,60,...,300} {
    \draw [dashed,thick] ({cos(\t)},{sin(\t)},0) -- ({cos(\t+60)},{sin(\t+60)},0);
    \draw [dashed,thick] (0,0,0) -- ({cos(\t-180)},{sin(\t-180)},0);
    }
  \foreach \t in {240,180,300,120} {
    \tdplottransformmainscreen{cos(\t)}{sin(\t)}{0}
    \shadedraw[tdplot_screen_coords, ball color = cyan!40,opacity=0.7] (\tdplotresx,\tdplotresy) circle (0.5) node [opacity=1] at ({\f*\tdplotresx},{\f*\tdplotresy)},0) {\small #1};
    }
  \tdplottransformmainscreen{0}{0}{0}
  \shadedraw[tdplot_screen_coords, ball color = #3, opacity=0.9] (\tdplotresx,\tdplotresy) circle (0.5) node [opacity=1,above] at (0,0) {\small #2};
  \foreach \t in {0,60} {
    \tdplottransformmainscreen{cos(\t)}{sin(\t)}{0}
    \shadedraw[tdplot_screen_coords, ball color = cyan!40,opacity=0.7] (\tdplotresx,\tdplotresy) circle (0.5) node [opacity=1] at ({\f*\tdplotresx},{\f*\tdplotresy)},0) {\small #1};
    }
  }}}
  \tikzset{pics/B/.style n args={2}{code={
  \foreach \t in {0,60,...,300} {
    \draw [dashed,thick] ({cos(\t)},{sin(\t)},0) -- ({cos(\t+60)},{sin(\t+60)},0);
    \draw [dashed,thick] (0,0,0) -- ({cos(\t-180)},{sin(\t-180)},0);
    }
  \foreach \t in {150,270} {
    \tdplottransformmainscreen{2/sqrt(3)*cos(\t)}{2/sqrt(3)*sin(\t)}{0}
    \shadedraw[tdplot_screen_coords, ball color = magenta!40,opacity=0.7] (\tdplotresx,\tdplotresy) circle (0.5) node [opacity=1]  at ({\f*\tdplotresx},{\f*\tdplotresy)},0) {\small #2};
    }
  \foreach \t in {210,330,90} {
    \tdplottransformmainscreen{1/sqrt(3)*cos(\t)}{1/sqrt(3)*sin(\t)}{0}
    \shadedraw[tdplot_screen_coords, ball color = cyan!40,opacity=0.7] (\tdplotresx,\tdplotresy) circle (0.5) node [opacity=1] at ({\f*\tdplotresx},{\f*\tdplotresy)},0) {\small #1};
    }
  \foreach \t in {30} {
    \tdplottransformmainscreen{2/sqrt(3)*cos(\t)}{2/sqrt(3)*sin(\t)}{0}
    \shadedraw[tdplot_screen_coords, ball color = magenta!40,opacity=0.7] (\tdplotresx,\tdplotresy) circle (0.5) node [opacity=1] at ({\f*\tdplotresx},{\f*\tdplotresy)},0) {\small #2};
    }
  }}}
  \tikzset{pics/C/.style n args={2}{code={
  \foreach \t in {0,60,...,300} {
    \draw [dashed,thick] ({cos(\t)},{sin(\t)},0) -- ({cos(\t+60)},{sin(\t+60)},0);
    \draw [dashed,thick] (0,0,0) -- ({cos(\t-180)},{sin(\t-180)},0);
    }
  \foreach \t in {210} {
    \tdplottransformmainscreen{2/sqrt(3)*cos(\t)}{2/sqrt(3)*sin(\t)}{0}
    \shadedraw[tdplot_screen_coords, ball color = magenta!40,opacity=0.7] (\tdplotresx,\tdplotresy) circle (0.5) node [opacity=1] at ({\f*\tdplotresx},{\f*\tdplotresy)},0) {\small #1};
    }
  \foreach \t in {150,270,30} {
    \tdplottransformmainscreen{1/sqrt(3)*cos(\t)}{1/sqrt(3)*sin(\t)}{0}
    \shadedraw[tdplot_screen_coords, ball color = cyan!40,opacity=0.7] (\tdplotresx,\tdplotresy) circle (0.5) node [opacity=1] at ({\f*\tdplotresx},{\f*\tdplotresy)},0) {\small #1};
    }
  \foreach \t in {330,90} {
    \tdplottransformmainscreen{2/sqrt(3)*cos(\t)}{2/sqrt(3)*sin(\t)}{0}
    \shadedraw[tdplot_screen_coords, ball color = magenta!40,opacity=0.7] (\tdplotresx,\tdplotresy) circle (0.5) node [opacity=1] at ({\f*\tdplotresx},{\f*\tdplotresy)},0) {\small #2};
    }
  }}}

\pic [transform shape] at (0,0,{-\d}) {B={1}{2}};
\pic [transform shape] at (0,0,0) {A={1}{0}{blue!30}};
\pic [transform shape] at (0,0,{\d}) {C={1}{2}};

\foreach \t in {30,150,270} {\draw [thick,dotted] ({1/sqrt(3)*cos(\t)},{1/sqrt(3)*sin(\t)},0) --++ (0,0,\d);};
\foreach \t in {150,270,30} {\draw [thick,dotted] ({1/sqrt(3)*cos(\t)},{1/sqrt(3)*sin(\t)},0) --++ (0,0,{-\d});};

\draw [very thick,>=stealth,->] (0,0,0) -- (1,0,0) node [above,xshift=-3pt] {$\bm{a}$};
\draw [very thick,>=stealth,->] (0,0,0) -- ({cos(120)},{sin(120)},0) node [above] {$\bm{b}$};

\node at (2,0,{1.75*\d}) {\large a)};

\begin{scope}[xshift=4.5cm]
    \pic [transform shape] at (0,0,{-\d}) {A={}{0}{blue!30}};
    \pic [transform shape] at (0,0,0) {B={}{}};
    \pic [transform shape] at (0,0,{\d}) {A={}{3}{purple!70}};

    \foreach \t in {30,150,270} {\draw [thick,dotted] ({1/sqrt(3)*cos(\t)},{1/sqrt(3)*sin(\t)},0) --++ (0,0,\d);};
    \foreach \t in {150,270,30} {\draw [thick,dotted] ({1/sqrt(3)*cos(\t)},{1/sqrt(3)*sin(\t)},0) --++ (0,0,{-\d});};

    \node at (2,0,{1.75*\d}) {\large b)};
\end{scope}

\begin{scope}[xshift=9cm]
    \pic [transform shape] at (0,0,{-\d}) {A={}{0}{blue!30}};
    \pic [transform shape] at (0,0,0) {B={}{}};
    \pic [transform shape] at (0,0,{\d}) {C={}{}};
    
    \foreach \t in {30,150,270} {\draw [thick,dotted] ({1/sqrt(3)*cos(\t)},{1/sqrt(3)*sin(\t)},0) --++ (0,0,\d);};
    \foreach \t in {150,270,30} {\draw [thick,dotted] ({1/sqrt(3)*cos(\t)},{1/sqrt(3)*sin(\t)},0) --++ (0,0,{-\d});};

    \node at (2,0,{1.75*\d}) {\large c)};
\end{scope}

\end{tikzpicture}
    \caption{a) Local atomic environment in the Barlow sequence $(\cdots BAC\cdots)$. The layers are offset along the $\hat{\bm{c}}$ axis for visual clarity. The reference atom is marked ~``0'' and is shown in blue. The sets of all~1nn and~2nn atoms are marked~``1'' and~``2'', shown in cyan and magenta respectively. b)~Atomic environment in the sequence $(\cdots ABA\cdots)$. Because the corresponding H\"{a}gg code, $(\cdots+-\cdots)$, now contains the substring $(+-)$, the ``$0$'' atom has a~3nn in the second layer above (marked~``3''). c)~Atomic environment in the sequence $(\cdots ABC\cdots)$. The ``0'' atom now has no~3nn [up to the distance cutoff considered, see the discussion below Eq.~(\ref{eq:close_packed_chi_definition})], since the H\"{a}gg code entries do not change sign.}
    \label{fig:env}
\end{figure*}
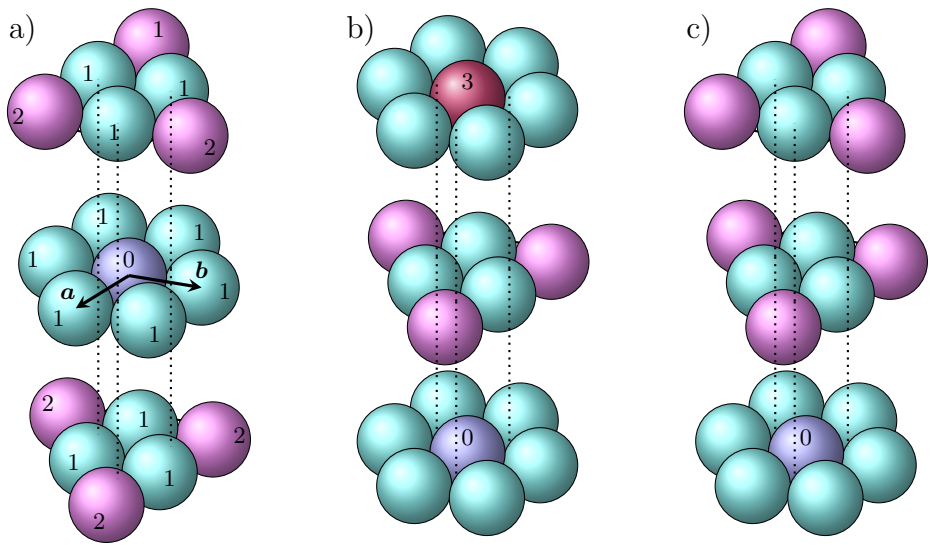

\subsection{\label{ssec:close_packed_mapping}Mapping onto the disordered chain}

We now show that the model of a disordered 1D chain in Sec.~\ref{sec:disordered_chain} may be adapted to describe close-packed solids. First, we introduce a generalization of Eq.~(\ref{eq:chain_hamiltonian}) that includes occurrences of the second-neighbour hopping with amplitude $t'$:
\begin{equation}
    \hat{H}=\begin{pmatrix}
    \varepsilon_0 & -te^{i\phi_1} & W' &  &   &   \\[0.3em]
    -te^{-i\phi_1} & \varepsilon_0 & \ddots & \ddots &   &   \\[0.3em]
    W' & \ddots & \ddots & \ddots & \ddots &   \\[0.3em]
    & \ddots & \ddots & \ddots & \ddots & \ddots \\[0.3em]
\end{pmatrix}.
\label{eq:close_packed_hamiltonian_1D_chain}
\end{equation}
In the same vein as $\Wk{m}$, the matrix element $W'$ is equal to $-t'$ wherever a second-neighbour hopping process occurs; otherwise, it is equal to zero. The analytic results derived in Sec.~\ref{sec:disordered_chain} hold only when all of the $W'$ elements vanish. In Eq.~(\ref{eq:close_packed_hamiltonian_1D_chain}), we represent the Hamiltonian as a semi-infinite matrix, anticipating its later use to describe a semi-infinite lead in transport calculations.

Motivated by the property (\ref{eq:close_packed_Vk_complex_conjugate}), we introduce the shorthand notation
\begin{equation}
    \lvert V_{\kp}^{-\sigma_m}\rvert
    =\lvert \Vk{m}\rvert
    \equiv V_{\kp}.
    \label{eq:close_packed_mod_Vk_definition}
\end{equation}
It now follows that the Hamiltonian of the phase-disordered chain, Eq.~(\ref{eq:chain_hamiltonian}), is the same as the Bloch Hamiltonian of a close-packed solid, Eq.~(\ref{eq:close_packed_hamiltonian}), with the $W$ terms set to zero. This can be seen by making the transcriptions
\begin{equation}
    \varepsilon_0\leftrightarrow \Ek,\ \
    t\leftrightarrow V_{\kp},\ \
    t'\leftrightarrow t_3,\ \
    \phi_{m}\leftrightarrow\arg{\Vk{m}}\pm\pi.
    \label{eq:close_packed_transcriptions}
\end{equation}
The $\kp$-dependence of $\Ek$ is the reason for retaining $\varepsilon_0$ back in Eq.~(\ref{eq:chain_hamiltonian}). The 1nn and~2nn hoppings in the 3D close-packed structure ($t_1$ and $t_2$) contribute to the nearest-neighbour hopping $(t)$ in the effective 1D chain. The role of the second-neighbour hopping in the 1D chain ($t'$) is played by the 3nn hoppings in the close-packed solid ($t_3$).

As an example of the utility of Eq.~(\ref{eq:close_packed_transcriptions}), we consider a situation where the 3nn hopping is negligible, i.e., $t_3=0$. Energies of the electronic states in an $N$-layer close-packed solid can be obtained from Eq.~(\ref{eq:chain_eigenvalues}),
\begin{equation}
    \epsilon_{\kp,j}=\Ek+2V_{\kp}\cos{\beta_j},
    \label{eq:close_packed_energy_levels}
\end{equation}
where $\kp$ is the momentum perpendicular to the stacking direction and $\beta_j$ is given by Eq. (\ref{eq:chain_wavevector}). Since the right-hand side of Eq.~(\ref{eq:close_packed_energy_levels}) is independent of the $\sigma$'s, we arrive at the result that the electronic energy levels in a close-packed solid are completely insensitive to the stacking sequence, as long as the hoppings to~3nn and beyond are negligible~\cite{he2025}. 
In the thermodynamic limit ($N\to\infty$), Eq. (\ref{eq:close_packed_energy_levels}) takes the form
\begin{equation}
    \epsilon_{\kp}(\beta)=\Ek+2V_{\kp}\cos{\beta},
    \label{eq:close_packed_energy_levels-infinite-N}
\end{equation}
where $\beta\in (0,\pi)$ is a continuous parameter. It follows from Eq. (\ref{eq:close_packed_energy_levels-infinite-N}) that the energy levels take values between
\begin{equation}
    E_{\mathrm{min}}=\min_{\kp}\,(\varepsilon_{\kp}-2V_{\kp}),\quad
    E_{\mathrm{max}}=\max_{\kp}\,(\varepsilon_{\kp}+2V_{\kp}),
\label{eq:close_packed_band_edges}
\end{equation}
which define the lower and upper band edges, respectively.

\section{\label{sec:stacking_fault_transmission}Stacking fault transmission}

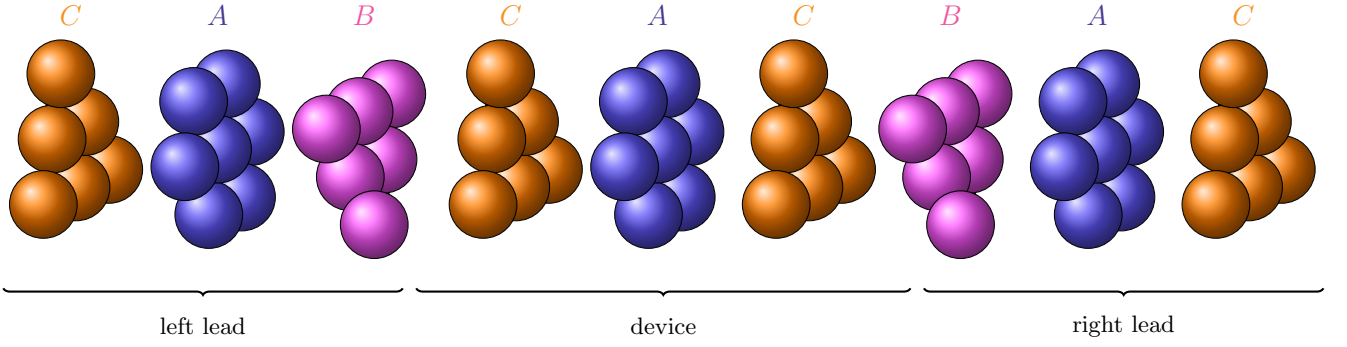
\begin{figure*}[tb]
    \centering
    \tdplotsetmaincoords{60}{135}     
\begin{tikzpicture}[tdplot_main_coords,scale=0.9,rotate=-90]
  \def\d{2.5}
  \def\f{1.32}


  \tikzset{A/.pic={
  \foreach \t in {0,60,...,300} {
    \draw [dashed,thick] ({cos(\t)},{sin(\t)},0) -- ({cos(\t+60)},{sin(\t+60)},0);
    \draw [dashed,thick] (0,0,0) -- ({cos(\t-180)},{sin(\t-180)},0);
    }
  \foreach \t in {240,180,300,120} {
    \tdplottransformmainscreen{cos(\t)}{sin(\t)}{0}
    \shadedraw[tdplot_screen_coords, ball color = BlueViolet!70,opacity=0.9] (\tdplotresx,\tdplotresy) circle (0.5) node [opacity=1] at ({\f*\tdplotresx},{\f*\tdplotresy)},0) {};
    }
  \node [color=BlueViolet,rotate=90] at (1.4,-1.4) {\large $A$};
  \tdplottransformmainscreen{0}{0}{0}
  \shadedraw[tdplot_screen_coords, ball color = BlueViolet!70,opacity=0.9] (\tdplotresx,\tdplotresy) circle (0.5) node [opacity=1,above] at (0,0) {};
  \foreach \t in {0,60} {
    \tdplottransformmainscreen{cos(\t)}{sin(\t)}{0}
    \shadedraw[tdplot_screen_coords, ball color = BlueViolet!70,opacity=0.9] (\tdplotresx,\tdplotresy) circle (0.5) node [opacity=1] at ({\f*\tdplotresx},{\f*\tdplotresy)},0) {};
    }
  }}
  \tikzset{B/.pic={
  \foreach \t in {0,60,...,300} {
    \draw [dashed,thick] ({cos(\t)},{sin(\t)},0) -- ({cos(\t+60)},{sin(\t+60)},0);
    \draw [dashed,thick] (0,0,0) -- ({cos(\t-180)},{sin(\t-180)},0);
    }
  \foreach \t in {270} {
    \tdplottransformmainscreen{2/sqrt(3)*cos(\t)}{2/sqrt(3)*sin(\t)}{0}
    \shadedraw[tdplot_screen_coords, ball color = Rhodamine!80,opacity=0.9] (\tdplotresx,\tdplotresy) circle (0.5) node [opacity=1]  at ({\f*\tdplotresx},{\f*\tdplotresy)},0) {};
    }
  \node [color=Rhodamine,rotate=90] at (1.4,-1.4) {\large $B$};
  \foreach \t in {210,330,90} {
    \tdplottransformmainscreen{1/sqrt(3)*cos(\t)}{1/sqrt(3)*sin(\t)}{0}
    \shadedraw[tdplot_screen_coords, ball color = Rhodamine!80,opacity=0.9] (\tdplotresx,\tdplotresy) circle (0.5) node [opacity=1] at ({\f*\tdplotresx},{\f*\tdplotresy)},0) {};
    }
  \foreach \t in {30,150} {
    \tdplottransformmainscreen{2/sqrt(3)*cos(\t)}{2/sqrt(3)*sin(\t)}{0}
    \shadedraw[tdplot_screen_coords, ball color = Rhodamine!80,opacity=0.9] (\tdplotresx,\tdplotresy) circle (0.5) node [opacity=1] at ({\f*\tdplotresx},{\f*\tdplotresy)},0) {};
    }
  }}
  \tikzset{C/.pic={
  \foreach \t in {0,60,...,300} {
    \draw [dashed,thick] ({cos(\t)},{sin(\t)},0) -- ({cos(\t+60)},{sin(\t+60)},0);
    \draw [dashed,thick] (0,0,0) -- ({cos(\t-180)},{sin(\t-180)},0);
    }
  \foreach \t in {210} {
    \tdplottransformmainscreen{2/sqrt(3)*cos(\t)}{2/sqrt(3)*sin(\t)}{0}
    \shadedraw[tdplot_screen_coords, ball color = BurntOrange,opacity=0.9] (\tdplotresx,\tdplotresy) circle (0.5) node [opacity=1] at ({\f*\tdplotresx},{\f*\tdplotresy)},0) {};
    }
  \node [color=BurntOrange,rotate=90] at (1.4,-1.4) {\large $C$};
  \foreach \t in {150,270,30} {
    \tdplottransformmainscreen{1/sqrt(3)*cos(\t)}{1/sqrt(3)*sin(\t)}{0}
    \shadedraw[tdplot_screen_coords, ball color = BurntOrange,opacity=0.9] (\tdplotresx,\tdplotresy) circle (0.5) node [opacity=1] at ({\f*\tdplotresx},{\f*\tdplotresy)},0) {};
    }
  \foreach \t in {330,90} {
    \tdplottransformmainscreen{2/sqrt(3)*cos(\t)}{2/sqrt(3)*sin(\t)}{0}
    \shadedraw[tdplot_screen_coords, ball color = BurntOrange,opacity=0.9] (\tdplotresx,\tdplotresy) circle (0.5) node [opacity=1] at ({\f*\tdplotresx},{\f*\tdplotresy)},0) {};
    }
  }}

\path [transform shape] (0,0,{0*\d}) pic {C};
\path [transform shape] (0,0,{1*\d}) pic {A};
\path [transform shape] (0,0,{2*\d}) pic {B};

\path [transform shape] (0,0,{3*\d}) pic {C};
\path [transform shape] (0,0,{4*\d}) pic {A};
\path [transform shape] (0,0,{5*\d}) pic {C};

\path [transform shape] (0,0,{6*\d}) pic {B};
\path [transform shape] (0,0,{7*\d}) pic {A};
\path [transform shape] (0,0,{8*\d}) pic {C};

\draw [tdplot_screen_coords,thick,decoration={brace,mirror,raise=0.5cm},decorate] (1.5,{-1}) -- (1.5,{2*\d-0.1}) node [midway,shift={(0,-1)}] {left lead};

\draw [tdplot_screen_coords,thick,decoration={brace,mirror,raise=0.5cm},decorate] (1.5,{2*\d+0.1}) -- (1.5,{5*\d-0.1}) node [midway,shift={(0,-1)}] {device};

\draw [tdplot_screen_coords,thick,decoration={brace,mirror,raise=0.5cm},decorate] (1.5,{5*\d+0.1}) -- (1.5,{7*\d+1}) node [midway,shift={(0,-1)}] {right lead};



\end{tikzpicture}
    \caption{3D rendering of the stacking fault described by Eqs. (\ref{eq:stacking_fault_barlow_sequence}) and (\ref{eq:stacking_fault_hagg_code}). The left and right leads consist of repeating $(CAB)$ and $(BAC)$ units, respectively, while the device is taken as the three-layer sequence $(CAC)$. Note that the only~3nn hopping process occurs between the $C$-layers in the device, i.e., for every atom in the left $C$-layer in the device, there is a corresponding atom in the right $C$-layer at a displacement of $\sqrt{8/3}~\bm{\hat{z}}$.}
    \label{fig:stacking_fault_3D}
\end{figure*}

In this section, we combine the results of Secs.~\ref{sec:disordered_chain} and~\ref{sec:close_packed_solids} to derive a formula for the transmission function of a single stacking fault in an fcc arrangement. We choose an fcc background as it is naturally devoid of 3nn hopping processes, as shown in Fig. \ref{fig:env}c. The H\"agg code representation has all entries as `$+$' (or equivalently, all entries as `$-$'). There are no pairs of atoms at the 3nn distance, as there are no $\chi$ variables that are either $(+-)$ or $(-+)$.

We consider a single stacking fault in an otherwise well-ordered, infinite fcc solid, represented by
\begin{subequations}
    \begin{align}
        & \mbox{Barlow:}\quad
        \cdots CABCAB\lvert CAC \rvert BACBAC\cdots
        \label{eq:stacking_fault_barlow_sequence}, \\
        & \mbox{H\"{a}gg:}\quad
        \left(\cdots+++\right)+\left[+-\right]-\left(---\cdots\right).
        \label{eq:stacking_fault_hagg_code}
    \end{align}
\end{subequations}
In Eq.~(\ref{eq:stacking_fault_barlow_sequence}), the two semi-infinite ``leads'' are shown separated by vertical bars. The left and right leads consist of repeating three-letter sequences $CAB$ and $BAC$ respectively, representing two distinct fcc arrangements. The leads are separated by a three-layer ``device'', described by $CAC$. The 3D picture of this configuration is shown in  Fig.~\ref{fig:stacking_fault_3D}.

The layer configuration is expressed as a H\"{a}gg code in Eq.~(\ref{eq:stacking_fault_hagg_code}). 
Chirality variables within each lead are enclosed by parentheses, while those internal to the device are enclosed by square brackets. The~`$+$' and~`$-$' signs that are not enclosed within brackets show the coupling between each lead and the device, e.g., the left lead couples to the device via the positive-chirality matrix element $V_{\kp}^{+}$. We choose the device to consist of three layers in this way so that (i) 3nn hopping occurs entirely within the device region and (ii) there is no direct hopping between left and right leads. 

\begin{figure}[t]
    \centering
    \begin{tikzpicture}
    \def\a{1.1}
    \foreach \i in {1,...,3} {\draw (\a*\i,0) circle (0.2) node {$\i$};}
    \node at ({-1*\a},0) {$\cdots$};
    \node at ({5*\a},0) {$\cdots$};
    \foreach \i in {0,4} {\fill (\a*\i,0) circle (0.2) node [color=white] {$\i$};}
    \draw [gray,->,>=stealth,very thin] ({0*\a+0.1},0.3) to [out=45,in=135] node [midway, above] {$-te^{i\phi_0}$} ({1*\a-0.1},0.3);
    \draw [gray,->,>=stealth,very thin] ({1*\a+0.1},0.3) to [out=45,in=135] node [midway, above] {$-te^{i\phi_1}$} ({2*\a-0.1},0.3);
    \draw [gray,->,>=stealth,very thin] ({2*\a+0.1},0.3) to [out=45,in=135] node [midway, above] {$-te^{i\phi_2}$} ({3*\a-0.1},0.3);
    \draw [gray,->,>=stealth,very thin] ({3*\a+0.1},0.3) to [out=45,in=135] node [midway, above] {$-te^{i\phi_3}$} ({4*\a-0.1},0.3);
    \draw [gray,->,>=stealth,very thin] ({1*\a},-0.3) --++ (0,-0.4) --++ ({2*\a},0) node [fill=white,midway] {$-t'e^{i\alpha}$} --++ (0,0.4);
\end{tikzpicture}
    \caption{Lead-device-lead arrangement for transmission through a stacking fault. The device and lead sites are represented by open and closed circles, respectively. First-neighbour hopping with phases $\phi_m$ exist between all sites in the system, while a single second-neighbour hopping process acts between the endpoints of the device.}
    \label{fig:stacking_fault}
\end{figure}
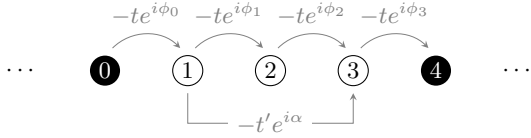

In the remainder of this section, we focus on a single in-plane momentum $\kp$. For ease of notation, we formulate the problem in the language of a disordered 1D chain, which is equivalent to a close-packed solid as shown in Sec.~\ref{ssec:close_packed_mapping}. The setup is as depicted in Fig.~\ref{fig:stacking_fault}. The phases $\phi_m$ shown there arise from the phases of $V_{\kp}^{\sigma_m}$. We solve a more general problem where the phases of $V_{\kp}^{\pm}$ take arbitrary values within the device. We also allow the second nearest-neighbour hopping process to carry a phase $\alpha$, with $\alpha$ being zero in the physically relevant case where $t'$ arises from the 3nn hoppings described by Eq.~(\ref{eq:Wk}).
We opt not to gauge away any of these phases, in order to clearly see the role they play. We do, however, choose a basis in which all the $\phi$'s have been eliminated from the lead Hamiltonians and the lead-device couplings. The existence of such a basis is demonstrated in Appendix~\ref{app:gauge_transformation}. 

The Hamiltonian corresponding to the configuration described by Eq.~(\ref{eq:stacking_fault_hagg_code}) may be written in block form as
\begin{equation}
    \hat{H}=\begin{pmatrix}
        \hat{H}_L&\hat{\tau}_L&0\\[0.3em]
        \hat{\tau}_L^\dag&\hat{H}_D&\hat{\tau}_R^\dag\\[0.3em]
        0&\hat{\tau}_R&\hat{H}_R
    \end{pmatrix},
    \label{eq:stacking_fault_block_hamiltonian}
\end{equation}
where $\hat H_L$, $\hat H_R$, and $\hat H_D$ are the Hamiltonians of the left lead, right lead, and device respectively, all in isolation. 
For definiteness, we assume that both leads contain $N$ atoms and take the $N\rightarrow\infty$ limit eventually. Both $\hat H_L$ and $\hat H_R$ are $N\times N$ Toeplitz matrices having the form of Eq.~(\ref{eq:chain_hamiltonian}) (with all $\phi$'s equal to zero). The device Hamiltonian, $\hat{H}_D$, is a $3\times 3$ matrix given by
\begin{equation}
    \hat H_D=
    \begin{pmatrix}
        \varepsilon_0&-te^{i\phi_{1}}&-t'e^{i\alpha}\\[0.3em]
        -te^{-i\phi_{1}}&\varepsilon_0&-te^{i\phi_{2}}\\[0.3em]
        -t'e^{-i\alpha}&-te^{-i\phi_2}&\varepsilon_0
    \end{pmatrix},
    \label{eq:stacking_fault_device_hamiltonian}
\end{equation}
where the phases $\phi$ are defined by Eq.~(\ref{eq:close_packed_transcriptions}), with $\phi_1$ ($\phi_2$) corresponding to positive (negative) chirality, in accordance with Eq.~(\ref{eq:stacking_fault_hagg_code}). Lastly, $\hat\tau_L$ and $\hat\tau_R$ are $N\times3$ matrices that encode lead-device coupling. They each have a single nonzero matrix element,
\begin{subequations}
    \begin{align}
        \tau_{L,mn}&=-t\delta_{m,N}\,\delta_{n,1},
        \label{eq:stacking_fault_tau_L}\\
        \tau_{R,mn}&=-t\delta_{m,1}\,\delta_{n,3},
        \label{eq:stacking_fault_tau_R}
    \end{align}
\end{subequations}
owing to the locality of the lead-device coupling.

The transmission function through the device is computed using the Caroli formula~\cite{meir_PRB_1992},
\begin{equation}
    T_{1\mathrm{D}}(E)=\mbox{Tr}\left[\hat G_D^\dagger(E)\,\hat\Gamma_R(E)\,\hat G_D(E)\,\hat \Gamma_L(E)\right].
    \label{eq:stacking_fault_Caroli_formula}
\end{equation}
The subscript `$1\mathrm{D}$' here emphasizes that the transmission function is calculated in an effective 1D chain model for a given $\kp$. For the sake of completeness, we provide a derivation of Eq.~(\ref{eq:stacking_fault_Caroli_formula}) in Appendix~\ref{app:Caroli}. 
We describe the terms in this expression below, using the subscript $a\in\{L,R\}$ to label the two leads. 
We use $\hat{g}_a(E)$ to denote the retarded Green's function of the $a^{\mathrm{th}}$ lead in isolation, as given by Eq.~(\ref{eq:chain_retarded_greens_function_with_phases}). We introduce the self-energy $\hat{\Sigma}_a(E)=\hat{\tau}_a^\dag\,\hat{g}_a(E)\,\hat{\tau}_a$, arising from the coupling of the $a^{\mathrm{th}}$ lead to the device. After performing the matrix multiplications, we find that $\Sigma_{L,mn}$ depends only on $g_{L,NN}$, which may be replaced with $g_{L,11}$, according to Eq.~(\ref{eq:chain_greens_function_symmetry}).
In terms of components, the self-energy matrices are given by
\begin{subequations}
    \begin{align}
        \Sigma_{L,mn}(E)
        &=te^{-i\theta}\,\delta_{m,1}\,\delta_{n,1},
        \label{eq:stacking_fault_sigma_L}\\
        \Sigma_{R,mn}(E)
        &=te^{-i\theta}\,\delta_{m,3}\,\delta_{n,3}.
        \label{eq:stacking_fault_sigma_R}
    \end{align}
\end{subequations}
Here, $\theta$ encodes the energy of an incoming electron as defined in Eq.~(\ref{eq:chain-energy-parametrization}). 
We also introduce the level-width matrices $\hat{\Gamma}_a(E)=-2\im{\hat{\Sigma}_a(E)}$ with the following elements:
\begin{subequations}
    \begin{align}
        \Gamma_{L,mn}(E)
        &=2t\sin{\theta}\,\delta_{m,1}\,\delta_{n,1},
        \label{eq:stacking_fault_gamma_L}\\
        \Gamma_{R,mn}(E)
        &=2t\sin{\theta}\,\delta_{m,3}\,\delta_{n,3}.
        \label{eq:stacking_fault_gamma_R}
    \end{align}
\end{subequations}
The Green's function of the device, $\hat{G}_D(E)$, including the lead-device couplings, is given by
\begin{equation}
    \hat{G}_D(E)=\left[E-\hat{H}_D-\hat{\Sigma}_L(E)-\hat{\Sigma}_R(E)\right]^{-1}.
    \label{eq:stacking_fault_device_Greens_function_definition}
\end{equation}

Substituting Eqs.~(\ref{eq:stacking_fault_gamma_L}) and~(\ref{eq:stacking_fault_gamma_R}) into Eq.~(\ref{eq:stacking_fault_Caroli_formula}), we obtain
\begin{equation}
    T_{1\mathrm{D}}(E)=4t^2\sin^2{\theta}\,\left| G_{D,31}(E)\right|^2.
    \label{eq:stacking_fault_Caroli_formula_simplified}
\end{equation}
We see that it is sufficient to know only one matrix element of $\hat{G}_D(E)$ in order to calculate the transmission function. This is straightforwardly done via a cofactor expansion of Eq.~(\ref{eq:stacking_fault_device_Greens_function_definition}) and we find
\begin{widetext}
\begin{equation}
    T_{1\mathrm{D}}(E)=\frac{\left(1-4\Delta\cos{\Phi}\cos{\theta}+4\Delta^2\cos^2{\theta}\right)\sin^2{\theta}}{\left(\Delta^2\cos{\theta}-\Delta\cos{\Phi}+\sin{2\theta}\sin{\theta}\right)^2+\cos^2{2\theta}\,\sin^2{\theta}},\quad
    \Delta=\frac{t'}{t},\quad
    \Phi=\phi_1+\phi_2-\alpha.
    \label{eq:stacking_fault_transmission_function}
\end{equation}  
\end{widetext}
The angle $\Phi$ represents the net phase accrued in traversing the closed loop in the device -- forward from left-to-right (from site~1 to site~3 via site~2) and backwards (from site~3 to site~1 directly), see Fig.~\ref{fig:stacking_fault}. It can be interpreted as the net ``magnetic flux'' enclosed at the device. 
The transmission function does not depend on the individual phases, $\phi_1$ and $\phi_2$, but only on the net flux through the device. As a consistency check, we note that $T_{1\mathrm{D}}(E)$ is at most unity for any values of $\Delta$, $\Phi$, and $E$.

As expected on physical grounds, Eq.~(\ref{eq:stacking_fault_transmission_function}) shows that the transmission function correctly reduces to unity when $\Delta=0$, i.e., when the 3nn hopping $t_3$ (and therefore $t'$ in the 1D chain model) is set to zero. 
For $\Delta\ll1$, we may replace Eq.~(\ref{eq:stacking_fault_transmission_function}) by
\begin{equation}
    T_{1\mathrm{D}}(E)\simeq 1-\frac{\Delta^2\cos^2{\Phi}}{\sin^2{\theta}},
    \label{eq:stacking_fault_transmission_function_asymptotic}
\end{equation}
when $\theta$ is in the vicinity of $\pi/2$, corresponding to the incident energy away from the band edges.

As seen from Eq.~(\ref{eq:stacking_fault_transmission_function_asymptotic}), the leading-order correction to the transmission function is quadratic in $\Delta$. That the linear term must vanish can be seen from the following argument: a hypothetical linear-order correction would have the form $\Delta f(E)$, where $f$ is some function. In order for $T_{1\mathrm{D}}(E)$ not to exceed unity at this order, we must have $\Delta f(E)\leq 0$ (since the zeroth-order part of Eq.~(\ref{eq:stacking_fault_transmission_function}) is already equal to one), irrespective of the sign of $\Delta$. The only way this can be ensured is when $f(E)=0$.

Lastly, it is interesting to note that, for $\Phi=0$, there exists a critical value of $\theta$, given by
\begin{equation}
    \sin\theta_c=\frac{\sqrt{4\Delta^2-1}}{2\Delta},
    \label{eq:stacking_fault_critical_angle}
\end{equation}
for which the numerator of Eq.~(\ref{eq:stacking_fault_transmission_function}) vanishes. This is analogous to total internal reflection of light rays propagating from a medium whose ``refractive index'' is equal to the denominator of Eq.~(\ref{eq:stacking_fault_critical_angle}) to another where it is equal to the numerator~\cite{sears1958}.

\section{\label{sec:random_chain}Random chain}

In this section, we consider a generalization of Eq.~(\ref{eq:stacking_fault_hagg_code}) where the device comprises many random chirality variables, as opposed to a single stacking fault. In H\"{a}gg code notation, we consider a system of the form
\begin{equation}
    \left(\cdots++\right)+\left[\sigma_1\sigma_2\cdots\sigma_{M-1}\right]+\left(++\cdots\right).
    \label{eq:random_chain_hagg_code}
\end{equation}
As in Sec.~\ref{sec:stacking_fault_transmission}, entries within parentheses in Eq.~(\ref{eq:random_chain_hagg_code}) represent the leads, while the square brackets represent the device. The unenclosed signs indicate the device-lead couplings. We make the following assumptions in order to avoid direct~3nn couplings between the device and either lead: we  fix the device-lead coupling to~`$+$'. We also fix the terminal entries in the device, $\sigma_1$ and $\sigma_{M-1}$, to be ~`$+$'. These assumptions allow us to use Eqs.~(\ref{eq:stacking_fault_tau_L}) and~(\ref{eq:stacking_fault_tau_R}) for the device-lead couplings.

The device is an $M$-layer stack, corresponding to a H\"{a}gg code of length $M-1$. Furthermore, we assume that $\sigma_2,\dots,\sigma_{M-2}$ are independent, Bernoulli-distributed random variables taking values~`$-$' and~`$+$' with probabilities $p$ and $q=1-p$, respectively. Qualitatively, $p$ is the probability of finding deviations from a uniform fcc arrangement.
From Eq.~(\ref{eq:close_packed_chi_definition}), a~3nn hopping appears between layers $m$ and $(m+2)$ of the device if the corresponding variable $\chi_m$ is equal to $(+-)$ or $(-+)$. According to our assumptions about the $\sigma_m$'s, each of these pairs occurs with likelihood $pq$. The probability of a 3nn coupling at a certain $m$ is then given by $P=2pq$. The ``maximally random'' configuration corresponds to choosing $p=0.5$.

The Hamiltonian corresponding to Eq.~(\ref{eq:random_chain_hagg_code}) is again of the block form of Eq.~(\ref{eq:stacking_fault_block_hamiltonian}). We adopt a basis in which all phases from the 1nn- and ~2nn-hoppings have been removed from the device, the leads as well as the lead-device couplings. The existence of such a basis is demonstrated in Appendix~\ref{app:gauge_transformation}. Since the device-lead couplings are of the same form as in Sec.~\ref{sec:stacking_fault_transmission}, the self-energies and level-width functions continue to be given by Eqs.~(\ref{eq:stacking_fault_sigma_L}) through~(\ref{eq:stacking_fault_gamma_R}). 

The device Hamiltonian, $\hat{H}_D$, is now of the form given in Eq.~(\ref{eq:close_packed_hamiltonian_1D_chain}). It contains a random arrangement of zero and nonzero matrix elements along its second off-diagonal, arising from the 3nn hoppings in the close-packed structure, with the nonzero entries equal to $-t'$. 
A direct calculation of $\hat{G}_D(E)$ via matrix inversion, as in Eq.~(\ref{eq:stacking_fault_device_Greens_function_definition}), is no longer possible. To calculate the transmission function, we therefore use a perturbative scheme developed in Ref.~\cite{gaoyang_PRB_2025}.

All information pertaining to the leads is incorporated into the self-energies and level-widths; once these are known, we no longer need the lead Hamiltonians or Green's functions. Thus, for ease of notation, we use $\hat{G}$ and $\hat{H}$ (without subscripts) to denote the device Green's function and Hamiltonian, respectively. We decompose $\hat{H}$ as  $\hat{H}=\hat{H}_0+\hat{V}$, where $\hat{H}_0$ is the tridiagonal Toeplitz part of $\hat{H}$ and $\hat{V}$ is a matrix that encodes the~3nn hoppings. The latter varies from one disorder realization to the next, with the nonzero entries given by
\begin{equation}
    V_{mn}=v_m\left(\delta_{n,m-2}+\delta_{n,m+2}\right),
    \label{eq:random_chain_V_matrix_elements}
\end{equation}
where $v_m=-t'$ if $\chi_m$ produces a~3nn hopping and is equal to zero otherwise.

Instead of Eq.~(\ref{eq:stacking_fault_Caroli_formula}), we obtain the following expression for the transmission function:
\begin{equation}
    T_{1\mathrm{D}}(E)=\mbox{Tr}\left[{\expval{\hat{G}^\dag \hat{\Gamma}_R\hat{G}}\hat{\Gamma}_L}\right],
    \label{eq:random_chain_averaged_Caroli_formula}
\end{equation}
where $\expval{\cdot}$ denotes an average over all possible disorder realizations, and $\hat{G}$ is the device Green's function given by 
\begin{equation}
    \hat{G}(E)=\bigl(E-\hat{H}_0-\hat{V}-\hat{\Sigma}_L-\hat{\Sigma}_R\bigr)^{-1}.
    \label{eq:random_chain_device_G}
\end{equation}
We can express $\hat{G}$ by the Dyson expansion
\begin{equation}
    \hat{G}=\hat{g}+\hat{g}\hat{V}\hat{g}+\hat{g}\hat{V}\hat{g}\hat{V}\hat{g}+\dots,
    \label{eq:random_chain_Dyson_equation}
\end{equation}
where $\hat{g}=\bigl(E-\hat{H}_0-\hat{\Sigma}_L-\hat{\Sigma}_R\bigr)^{-1}$ is the Green's function of the device without randomness. Using the explicit forms of the self-energies given by Eqs.~(\ref{eq:stacking_fault_sigma_L}) and~(\ref{eq:stacking_fault_sigma_R}), this latter quantity takes the form of the $M\times M$ matrix:
\begin{equation}
\hat{g}(E)=
\begin{pmatrix}
te^{i\theta} & t &   &        &        \\[0.3em]
t & 2t\cos{\theta} & t &        &        \\[0.3em]
  & t & \ddots & \ddots &   \\[0.3em]
  &   & \ddots & 2t\cos{\theta} & t \\[0.3em]
  &   &        & t      & te^{i\theta}
\end{pmatrix}^{\!\!-1}.
\label{eq:random_chain_device_greens_function_definition}
\end{equation} 
Here, the energy has been implicitly defined in terms of $\theta$, according to Eq.~(\ref{eq:chain-energy-parametrization}).
An elegant formula for the inverse of a tridiagonal matrix in the form of a three-term recurrence relation has been given in Ref.~\cite{usmani_1994}. As shown in Appendix~\ref{app:recursion}, when applied to Eq.~(\ref{eq:random_chain_device_greens_function_definition}), this formula yields
\begin{equation}
    g_{mn}(E)
    =\frac{(-1)^{m-n}}{2it}\,\frac{e^{-i|m-n|\theta}}{\sin{\theta}}.
    \label{eq:random_chain_unperturbed_device_greens_function}
\end{equation}

\begin{figure}[t]
    \centering
    \includegraphics{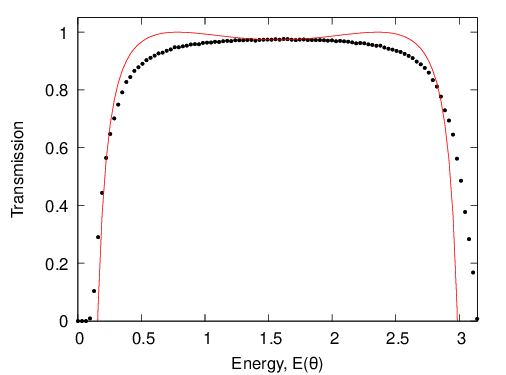}
    \caption{Transmission function vs the incident energy $E$ [the latter is parametrized as in Eq. (\ref{eq:chain-energy-parametrization})]. The solid circles were obtained by numerically calculating the device Green's function, inserting it into Eq.~(\ref{eq:random_chain_averaged_Caroli_formula}), and averaging over $10^3$ disorder realizations. The red line represents the quadratic correction, Eq.~(\ref{eq:random_chain_asymptotic_transmission_function}). We use the following values of the parameters: $\Delta=0.03$, $p=0.1$, and $M=200$.}
    \label{fig:quadratic_correction}
\end{figure}

The transmission function can be calculated to any desired order in $\hat{V}$ by substituting Eqs.~(\ref{eq:random_chain_Dyson_equation}) and~(\ref{eq:random_chain_unperturbed_device_greens_function}) into Eq.~(\ref{eq:random_chain_averaged_Caroli_formula}). The linear-order term vanishes identically, in line with the argument below Eq.~(\ref{eq:stacking_fault_transmission_function_asymptotic}). The quadratic-order correction contains sums involving the correlator
\begin{equation}
    \expval{v_mv_n}=\left(t^{\prime}\right)^2\times
    \left\{\begin{array}{ll}
        P,\quad & m=n,\\
        P^2,\quad & m\neq n.
    \end{array}\right.
    \label{eq:random_chain_two_point_correlator}
\end{equation}
With $m=n$, $\langle v_m^2\rangle$ is nonzero only if the corresponding $\chi$ variable allows a~3nn hopping, which occurs with probability $P$. If $m\neq n$, then the correlator decomposes into $\expval{v_m}\expval{v_n}$, since the $v$'s are independent random variables. We have $\expval{v_m}=\expval{v_n}=-t'P$.  
After a long, but straightforward calculation focusing on the $M\gg 1$ limit, we obtain the following result for the transmission function to leading order in $t'$:
\begin{equation}
    T_{1\mathrm{D}}(E)=1-\frac{MP(1-P)\Delta^2\cos^2{2\theta}}{\sin^2{\theta}},
\label{eq:random_chain_asymptotic_transmission_function}
\end{equation}
where $\Delta=t'/t$, as in Sec.~\ref{sec:stacking_fault_transmission}. In Fig.~\ref{fig:quadratic_correction}, we compare this leading-order analytical result to the transmission function obtained by numerically evaluating Eq.~(\ref{eq:random_chain_device_G}) and inserting it into Eq.~(\ref{eq:random_chain_averaged_Caroli_formula}). We see good qualitative agreement between the two approaches.

We note in passing the resemblance of Eq.~(\ref{eq:random_chain_asymptotic_transmission_function}) to Eq.~(\ref{eq:stacking_fault_transmission_function_asymptotic}). For $Mp\Delta^2\ll1$ and $E$ near the middle of the band, the randomly-stacked device can be viewed as a  single ``thick'' defect with an effective single-defect hopping that depends weakly on $E$.

The second term in Eq.~(\ref{eq:random_chain_asymptotic_transmission_function}) scales linearly with the number of layers in the device, which naively suggests that the transmission diverges towards $-\infty$ in the thermodynamic limit. 
However, as the transmission function has a lower bound of zero, we expect that the higher-order terms will cut off this divergence.
Based on the leading order result in Eq.~(\ref{eq:random_chain_asymptotic_transmission_function}), we conjecture that transmission function vanishes in the thermodynamic limit. We further conjecture that this holds true for any incident energy (i.e., for any $\theta$) and any disorder strength, in particular, for arbitrarily small values of $p$. 

\begin{figure*}
    \centering
    \includegraphics{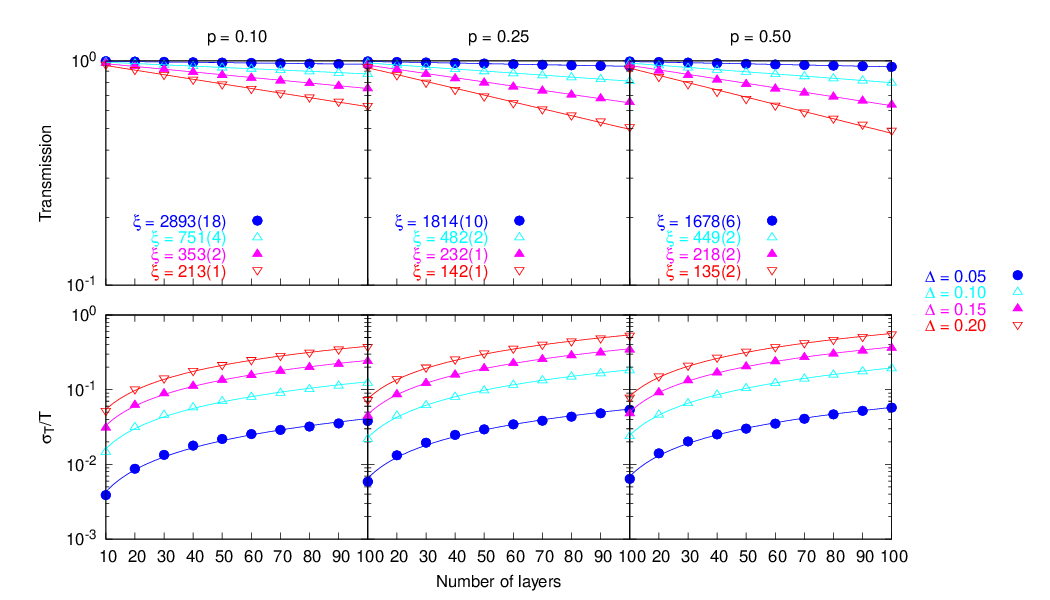}
    \caption{Top row: Transmission function (at the energy corresponding to $\theta=\pi/2$ and for various values of $p$), as determined by numerically calculating the device Green's function $\hat{G}$ and inserting it into Eq.~(\ref{eq:random_chain_averaged_Caroli_formula}). Each data point shown is the result of averaging over $10^5$ disorder realizations. The solid lines represent fits to $T=e^{-M/\xi}$. We note again that $p=0.5$ is the case of ``maximal randomness''. Bottom row: Ratio of the standard deviation of the transmission function to its average value. We observe that $T(E)$ is not a self-averaging quantity, in the sense that its fluctuations cannot be removed by sampling larger devices.}
    \label{fig:spread}
\end{figure*}

These conjectures are consistent with Anderson localization induced by disorder. The expression in Eq.~(\ref{eq:random_chain_asymptotic_transmission_function}) can be interpreted as the first-order term in the expansion of $T_{1\mathrm{D}}= e^{-M/\xi}$, which is the scaling form of conductivity in a 1D disordered electron system~\cite{Lee_TVR_1985}. 
This is supported by our numerical calculations shown in Fig.~\ref{fig:spread}. The figure plots the transmission function for devices with varying number of layers (various $M$ values). The transmission is obtained by numerically calculating the Green's function of Eq.~(\ref{eq:random_chain_device_G}) and inserting it into Eq.~(\ref{eq:random_chain_averaged_Caroli_formula}). The result is averaged over $10^5$ disorder realizations with fixed values of the parameters $t'/t$ and $p$. We find that the transmission function falls monotonically to zero as $M$ increases. We find the same qualitative behaviour regardless of the incident energy $E$ and the values of $t'/t$ and $p$.

As seen from Fig.~\ref{fig:spread}, the ratio of the standard deviation of the transmission to its average value does not decrease with the size of the disordered device. This indicates that the transmission function is not a self-averaging quantity. This behaviour is typical for 1D systems with quenched disorder, manifesting as universal conductance fluctuations in the metallic regime and a very broad (log-normal) distribution of the conductance in the localized regime~\cite{Beenakker1997,Muttalib2003}.

\section{\label{sec:momentum_average} Landauer resistance} 
\label{sec:Landauer}

The previous two sections describe the effects of phase disorder in a 1D chain. Here, we apply their results to the physical case of a stacking fault in a close-packed 3D solid. As described previously, every in-plane momentum $\kp$ serves as an independent channel for transport along the stacking direction.

The resistance $R$ of the system is related to the transmission properties via the Landauer formula~\cite{meir_PRB_1992}:
\begin{equation}
    R=\frac{h}{2e^2}\left[\sum_{i}T_i\right]^{-1}.
    \label{eq:momentum_averaged_landauer_buttiker}
\end{equation}
The sum over $i$ runs through all transport channels available to the system (a role played here by the in-plane momentum $\kp$). The transmission functions $T_i$ are to be evaluated at the Fermi energy $E_F$.

In Sec.~\ref{sec:stacking_fault_transmission}, we derived an explicit formula for the transmission function of a single stacking fault in the 1D model. The result, Eq.~(\ref{eq:stacking_fault_transmission_function}), contains the parameter $\Delta$, which, according to Eq.~(\ref{eq:close_packed_transcriptions}), is a function of the in-plane momentum $\kp$. It is also useful to introduce a channel-specific parametrization of the incident energy similar to Eq. (\ref{eq:chain-energy-parametrization}). For each value of $E$ satisfying $E_{\mathrm{min}}\leq E\leq E_{\mathrm{max}}$, the set of in-plane momenta within the first Brillouin zone participating in transmission is determined by the following condition:
\begin{equation}
    \Omega(E)=\{\kp:\Ek-2V_{\kp}\leq E\leq\Ek+2V_{\kp}\},
    \label{eq:T-k-integration-domain}
\end{equation}
which is obtained from Eq.~(\ref{eq:close_packed_energy_levels-infinite-N}). This condition suggests the following parametrization of $E$ in the channel $\kp$:
\begin{equation}
\label{eq:theta-parameter-k}
   E=\Ek+2V_{\kp}\cos\tk, 
\end{equation}
where $\tk\in(0,\pi)$. Thus, Eq.~(\ref{eq:stacking_fault_transmission_function}), with $\Delta\to\Delta_{\kp}=t_3/V_{\kp}$ and $\theta\to\theta_{\kp}$,
actually describes the transmission probability for a particular channel labelled by $\kp$, a quantity which we henceforth denote by $T(\kp,E)=T_{1\mathrm{D}}(E)$. 

We now introduce the overall transmission function of the sample, obtained by summing $T(\kp,E)$ over all $\kp$-channels:
\begin{equation}
    T(E)
    =\sum_{\kp}T(\kp,E)
    =A\displaystyle\int_{\Omega(E)}\frac{\di^2{\kp}}{(2\pi)^2}\:T(\kp,E),
    \label{eq:momentum_averaged_transmission}
\end{equation}
where $A$ is the cross-sectional area of the leads. The Landauer formula then takes the following form:
\begin{equation}
    R=\frac{h}{2e^2}\,\frac{1}{T(E_F)},
    \label{eq:momentum_average_resistance_def}
\end{equation}
where $T(E_F)$ is the overall transmission function evaluated at the Fermi energy.
We take two approaches to evaluating this last quantity. The first is fully numerical, where we determine the momentum domain $\Omega(E)$ and calculate the integral. In the second approach, we proceed analytically by assuming that $E_F$ lies close to the bottom of the energy band.

\subsection{Monte Carlo integration}

We define the resistance of a single unit cell as 
\begin{equation}
    R_{\mathrm{cell}}=R\,\mathcal{N}.
    \label{eq:momentum_averaged_Rcell}
\end{equation}
where $\mathcal{N}$ is the number of unit cells per layer. We determine the overall transmission function in Eq.~(\ref{eq:momentum_averaged_transmission}) using Monte Carlo integration. In order to do so, we require numerical values of the hopping parameters $t_1$, $t_2$, and $t_3$. We focus on the problem of fcc-stacked lithium with a single stacking fault. However, values for the hopping parameters are readily available only for the bcc structure~\cite{papaconstantopoulos_Springer_2014}. We use the same parameters, namely,
\begin{equation}
    t_1=1.01~\si{\electronvolt},\quad
    t_2=0.30~\si{\electronvolt},\quad
    t_3=-0.20~\si{\electronvolt},
    \label{eq:bcc_li_tight_binding_parameters}
\end{equation}
as ballpark estimates for the fcc structure. Assuming half-filling, the Fermi energy comes out to be $E_F=1.91~\si{\electronvolt}$.

\begin{figure}[t]
    \centering
    \includegraphics{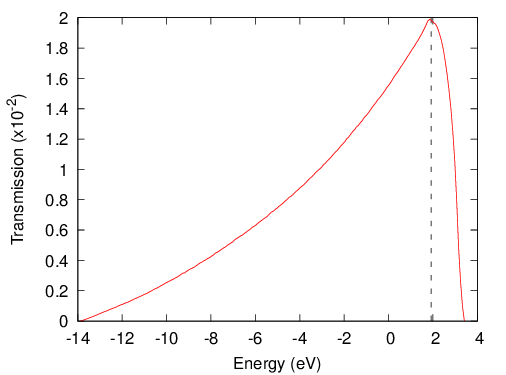}
    \caption{Transmission function of a single stacking fault in an fcc solid, given by Eq.~(\ref{eq:stacking_fault_transmission_function}) integrated over all $\kp\in\Omega(E)$. Transmission is plotted as a function of the incident energy $E$. The dashed line marks the location of the Fermi energy, $E_F=1.91~\si{\electronvolt}$. We estimate the error bars to be less than~5\% of the plotted transmission value.}
    \label{fig:integrated_transmission}
\end{figure}

Figure~\ref{fig:integrated_transmission} shows the result of integrating $T(\kp,E)$ over all $\kp\in\Omega(E)$. In the in-plane Brillouin zone, $5\times10^5$ points were sampled in the calculation of $T(E)$. At the Fermi energy, the overall transmission turns out to be $T(E_F)\simeq0.019$, which leads to the following estimate for the resistance of a single stacking fault:
\begin{equation}
    R_{\mathrm{cell}}
    \simeq0.66~\si{\mega\ohm}.
    \label{eq:momentum_averaged_stacking_fault_resistance}
\end{equation}
As shown in Appendix~\ref{app:BZ_integration}, this number translates to $r_S=0.092~\si{\mega\ohm\nano\meter\squared}$, where $r_S=RA$ is the specific stacking fault resistance. We have set the lattice constant to be $0.4~\si{\nano\meter}$ using the data for fcc lithium at $5~\si{\giga\pascal}$~\cite{Ackland2017}. 

Experimentally, resistance due to a single stacking-fault has been measured in graphite using carefully constructed pillars of stacked layers~\cite{Koren2014}. A similar measurement could, in principle, be performed for pressurized lithium or sodium, which are known to exhibit fcc structures~\cite{Ackland2017,Syassen2002}.
For Bernal-stacked graphite, $r_S=15(4)~\si{\mega\ohm\nano\meter\squared}$~\cite{Koren2014} is roughly two orders of magnitude larger than our result for the close-packed solids. We attribute this to the difference in the Fermi surface geometries: a smaller Fermi surface corresponds to fewer channels available for transport, and consequently, a larger resistance. In close-packed solids, there is a bulk Fermi surface at $E_F=1.91~\si{\electronvolt}$; on the other hand, Bernal graphite has a small Fermi energy $E_F=-22~\si{\milli\electronvolt}$~\cite{partoens_PRB_2006}, and a Fermi surface comprised of two small ``pockets'' around the Dirac points.

\subsection{Low-energy asymptotics}

When $E_F$ is close to the bottom of the energy band, it is possible to find an approximate parametrization of the integration domain $\Omega(E_F)$. 
To determine the location of the lower band edge, we observe that both~$\Ek$ and~$-2V_{\kp}$ take their minimum values of~$-6t_1$ and~$-6t_1-6t_2$, respectively, when~$\kp=0$ (assuming $t_1,t_2$ are positive). From Eq.~(\ref{eq:close_packed_band_edges}), it follows that~$E_{\mathrm{min}}=-12t_1-6t_2$.

For $E_F=E_{\mathrm{min}}$, the set $\Omega(E_F)$ shrinks to the single point $\kp=\bm{0}$. It follows by continuity that, for $E_F=E_{\mathrm{min}}+\delta E$, where $\delta E\ll|E_{\mathrm{min}}|$, the allowed values of $\kp$ will occupy a small neighbourhood surrounding $\kp=\bm{0}$. Thus, we expand Eq.~(\ref{eq:T-k-integration-domain}) to quadratic order in $\kp$. We obtain the equation of an ellipse, which is then mapped onto a disc by a rotation of the coordinates followed by a dilation. After introducing polar coordinates in the $\kp$-plane, the momentum integration domain defined by Eq.~(\ref{eq:T-k-integration-domain}) becomes
\begin{equation}
    k^2\leq\frac14\frac{\delta E}{t_1+t_2},
    \label{eq:momentum_average_fermi_surface_approximate}
\end{equation}
to leading order in $\delta E$.

Using Eq.~(\ref{eq:stacking_fault_transmission_function}), the transmission function for $E_F$ just above $E_{\mathrm{min}}$ is approximately given by
\begin{equation}
    T(\kp,E_F)\simeq\frac{\sin^2{\theta_{\kp}}}{\Dk^2}
    =\frac{V_{\kp}^2\,\sin^2\theta_{\kp}}{t_3^2},
    \label{eq:momentum_averaged_transmission_function_asymptotic}
\end{equation}
where we have made the mappings of Eq.~(\ref{eq:close_packed_transcriptions}), assuming that $t_3$ is nonzero and $\sin^2\theta_{\kp}\ll\Dk\ll1$. To see that~the inequality~$\Dk\ll1$ indeed holds, Eq.~(\ref{eq:Vk}) gives~$V_{\kp}\simeq3t_1+3t_2$ near~$\kp=0$. Thus,~$\Dk\simeq t_3/3t_1$, and using the estimates of  the hopping parameters given in Eq.~(\ref{eq:bcc_li_tight_binding_parameters}), we find $\Dk\simeq0.1$.

The overall transmission function, Eq.~(\ref{eq:momentum_averaged_transmission}), can now be evaluated by expanding Eq.~(\ref{eq:momentum_averaged_transmission_function_asymptotic}) about $\kp=0$ and integrating over the circular region described by Eq.~(\ref{eq:momentum_average_fermi_surface_approximate}). The result is, to leading order in $\delta E$,
\begin{equation}
    T(E_F)=\frac{3A}{16\pi a^2}\left(\frac{\delta E}{t_3}\right)^{\!2},
    \label{eq:overall_T}
\end{equation}
where $A$ is the cross-sectional area of the sample and $a$ is the lattice constant.
Using Eq.~(\ref{eq:overall_T}) in conjunction with the Landauer formula, as stated in Eqs.~(\ref{eq:momentum_average_resistance_def}) and~(\ref{eq:momentum_averaged_Rcell}), we estimate the resistance of a unit cell in the transverse direction to be
\begin{equation}
    R_{\mathrm{cell}}=(12.9~\si{\kilo\ohm})\times\frac{32\pi}{3\sqrt{3}}\left(\frac{t_3}{\delta E}\right)^{\!2}.
\end{equation}
We note that this estimate is not relevant to the physical case of lithium or sodium, where the Fermi energy is far from the bottom of the band. However, it may be applicable to artificial platforms such as topoelectric circuits~\cite{Thomale2018}, where transmission through close-packed structures can be simulated.  

\section{\label{sec:conclusions}Summary and conclusions}

We describe a framework for investigating electronic transport in close-packed solids. There are two key underlying assumptions in our analysis: (i) electrons reside in $s$-like valence orbitals, with hopping amplitudes that are direction-independent, and (ii) the hopping processes are short-ranged. Both assumptions can be reasonably made in the case of metallic lithium and sodium. They lead to structural frustration, with all close-packed structures being iso-energetic~\cite{he2025}. This picture can explain their seemingly random structures at low temperature and ambient pressure, where signatures of multiple close-packed structures are seen~\cite{Ackland2017}. 
Upon increasing pressure, lithium and sodium undergo a transition to a specific close-packed arrangement, corresponding to fcc structure~\cite{Ackland2017,Hanfland2002}. Our study can describe transport in the vicinity of this phase transition, where pressure can serve as a tuning parameter to control the density of stacking faults. A particularly exciting prospect is that of a pressure-tuned Anderson localization transition in mesoscopic lithium and sodium samples. As argued here, these 3D solids are effectively 1D systems, due to translational symmetry in the lateral directions. As a result, disorder may have an oversized impact and suppress transport altogether.

It was demonstrated in Ref.~\cite{he2025} that, under the above-mentioned assumptions, every close-packed structure has precisely the same electronic band structure. This is despite the fact that the structures themselves are not related by any physical symmetries, such as rotations or translations. Rather, they are related by a gauge symmetry that connects their respective Bloch Hamiltonians. In this article, we show that this gauge symmetry is fragile. We focus on electronic transport as a measurable quantity. With hoppings restrictred to first and second neighbours, transport is insensitive to the stacking sequence. However, a weak third-neighbour hopping suffices to differentiate between various stacking arrangements. As an illustration, we focus on fcc order and show that stacking faults lead to a discernible change in transmission. With no stacking faults, a sample of metallic lithium will ideally exhibit perfect transmission. When stacking faults appear, we will see a pronounced change in resistance. Indeed, as shown in Sec.~\ref{sec:random_chain}, when the density of stacking faults becomes sufficiently large, the Anderson localization will suppress transmission altogether.

A key feature of our approach is that we use the Landauer formalism for genuinely 3D solids. We exploit the stacked character of close-packed solids, which leads to lateral translation symmetry. This allows us to identify each value of lateral momentum as an independent channel for transport. Previous applications of the Landauer formula to bulk solids have been restricted to heuristic calculations~\cite{Koren2014} or certain regions of the Brillouin zone, e.g., in the vicinity of a Dirac point~\cite{Salehi2015,Tani2023}. Here, we calculate the exact transmission through a stacking fault in Sec.~\ref{sec:stacking_fault_transmission}, and through an arbitrary stacking sequence in Sec.~\ref{sec:random_chain}. These approaches can be extended to any stacked solid, e.g., to various arrangements of graphene layers.

Transport through stacking faults has been experimentally studied in graphite in Ref.~\cite{Koren2014}. This was achieved by carefully constructing micron-sized pillars and measuring resistivity along the pillar axis. The radius of pillars was designed to be less than the grain size so that the only relevant scattering centres are stacking faults along the axis. Variations in the stacking fault density gave rise to a broad width in measured resistance values. The specific resistance of a single stacking fault was extracted by fitting a statistical model for stacking fault distribution. A similar setup could be used for lithium and sodium to measure transport through stacking faults. Such measurements could shed light on the nature of the phase transition into the fcc structure, the energy cost of stacking faults and their distribution. We have presented estimates for the specific resistance due to a single stacking fault (see Sec.~\ref{sec:Landauer} and Appendix~\ref{app:stacking_fault}) which can help interpret such experiments.

\begin{acknowledgments}

This work was supported by the CGRS-D Scholarship 581595407 (CMW) and by Discovery Grants 2022-05240 (GR) and 2021-03705 (KS) from the Natural Sciences and Engineering Research Council of Canada. 

\end{acknowledgments}

\appendix

\section{\label{app:gauge_transformation}Gauge transformation}

\subsection{\label{app:gauge_transformation_isolated_H}Isolated Hamiltonian}

In this appendix, we show how the arbitrary phases in the Hamiltonian (\ref{eq:chain_hamiltonian}) can be removed by a unitary transformation~\cite{he2025}. We introduce the $N\times N$ diagonal matrix
\begin{equation}
    \hat{U}=\mbox{diag}\left(e^{i\varphi_1},e^{i\varphi_2},\cdots,e^{i\varphi_m},\cdots,e^{i\varphi_N}\right),
    \label{eq:gauge_transformation_U}
\end{equation}
and consider the matrix $\hat H'=\hat U\hat H\hat U^\dag$. Clearly, $\hat U$ does not alter the diagonal matrix elements of $\hat H$. On the other hand, the matrix elements along its off-diagonal transform according to
\begin{align}
    H'_{m,m+1}
    =-te^{i\left(\phi_{m}+\varphi_{m}-\varphi_{m+1}\right)}.
    \label{eq:gauge_transformation_H'}
\end{align}
We now choose the $\varphi$'s in such a way that the $\phi$'s do not appear in $\hat H'$. This constitutes a system of $N-1$ linear equations for the $N$ undetermined variables $\varphi_m$. By inspection of Eq.~(\ref{eq:gauge_transformation_H'}), the solution to this system is
\begin{equation}
    \varphi_m=\varphi_0+\sum_{r=1}^{m-1}\phi_{r},
    \label{eq:gauge_transformation_solution}
\end{equation}
where $\varphi_0$ is an arbitrary constant which we may freely set to zero here, but shall retain in Sec.~\ref{app:gauge_transformation_LDL}. 

If the second-nearest-neighbour hopping processes, with amplitude $t'$, are also present in $\hat H$, as in Eq.~(\ref{eq:close_packed_hamiltonian_1D_chain}) (physically corresponding to the 3nn hopping processes in a close-packed solid, with amplitude $t_3$), then the second-off-diagonal matrix elements transform according to
\begin{equation*}
    H'_{m,m+2}=-t'e^{i\left(\varphi_m-\varphi_{m+2}\right)}.
\end{equation*}
Recall from Eq.~(\ref{eq:close_packed_chi_definition}) that some of these matrix elements may vanish, if the corresponding $\chi_m$ is neither $(+-)$ nor $(-+)$. In the physical case, in which the $\phi_m$'s are determined from the argument of the matrix elements $\Vk{m}$, see Eq.~(\ref{eq:close_packed_transcriptions}), the second-off-diagonal elements are real. Indeed, it follows from Eq.~(\ref{eq:gauge_transformation_solution}) that
\begin{equation*}
    \varphi_{m}-\varphi_{m+2}
    =-\left(\phi_m+\phi_{m+1}\right).
\end{equation*}
This quantity is equal to zero, as the two $\phi$-phases are equal-and-opposite wherever $t'$ is nonzero. Therefore, in the physical case, the gauge transformation $\hat{U}$ removes all $\phi$-phases from the off-diagonal of $\hat{H}$, and at the same time introduces no new phases along the second off-diagonal.

\subsection{\label{app:gauge_transformation_LDL}Lead-device-lead arrangement}

The gauge transformation in Appendix \ref{app:gauge_transformation_isolated_H} was derived under the assumption that $\hat H$ is the Hamiltonian of an isolated~1D chain. We will now show how it may be generalized to accommodate a lead-device-lead arrangement. Suppose $\hat H$ is of the block form shown in Eq.~(\ref{eq:stacking_fault_block_hamiltonian}); that is, $\hat{H}$ is now the Hamiltonian of the entire lead-device-lead system. Then, under a unitary transformation with $\hat U=\hat{U}_L\oplus \hat{U}_D\oplus \hat{U}_R$, the matrix $\hat H$ becomes
\begin{equation}
    \hat H'=
    \begin{pmatrix}
        \hat{U}_L\hat{H}_L\hat{U}_L^\dag
        & \hat{U}_L\hat{\tau}_L\hat{U}_D^\dag
        & 0\\[0.3em]
        \hat{U}_D\hat{\tau}_L^\dag \hat{U}_L^\dag 
        & \hat{U}_D\hat{H}_D\hat{U}_D^\dag
        & \hat{U}_D\hat{\tau}_R^\dag \hat{U}_R^\dag\\[0.3em]
        0
        & \hat{U}_R\hat{\tau}_R\hat{U}_D^\dag
        & \hat{U}_R\hat{H}_R\hat{U}_R^\dag
    \end{pmatrix}.
    \label{eq:Hamiltonian}
\end{equation}
We take each of $\hat{U}_L,\hat{U}_D,\hat{U}_R$ to be matrices of the form shown in Eq.~(\ref{eq:gauge_transformation_U}). This generates three sets of undetermined phases $\varphi$, one from each subspace, which we denote by by $\varphi_{L,m}$, $\varphi_{D,m}$, and $\varphi_{R,m}$. We choose all three sets according to the solution given in Eq.~(\ref{eq:gauge_transformation_solution}). As shown in the previous section, this eliminates the $\phi$-phases from each diagonal sub-block of $\hat{H}'$, while still leaving three arbitrary phases $\varphi_{L,0}$, $\varphi_{D,0}$, and $\varphi_{R,0}$. Assuming that the lead-device coupling is local, and that only the $(N,1)$ element of $\hat{\tau}_L$ is nonzero, as in~Eq.~(\ref{eq:stacking_fault_tau_L}), one of these $\varphi_0$'s can be chosen to eliminate the leftover $\phi$-phase in $\hat{U}_L\hat{\tau}_L\hat{U}_D^\dag$, yielding
\begin{equation*}
    [\hat{U}_L\hat{\tau}_L\hat{U}_D^\dag]_{mn}=-t\delta_{m,N}\,\delta_{n,1}.
\end{equation*}
Similarly, a second $\varphi_0$ can be used to eliminate the $\phi$-phase from the $(1,N)$ element of $\hat{U}_R\hat{\tau}_R\hat{U}_D^\dag$. In this way, we have successfully gauged away all $\phi$-phases from $\hat{H}$, and, with no loss of generality, we can set the one remaining $\varphi_0$ to zero.

We may also consider a more restricted class of the gauge transformations, under which the $\phi$-phases in the device are left untouched. Namely, if $\hat{U}=\hat{U}_L\oplus\hat{1}_D\oplus \hat{U}_R$, where $\hat{1}_D$ is the identity matrix in the device subspace, then, by choosing $\hat{U}_L$ and $\hat{U}_R$ as above, all $\phi$'s can be removed from $\hat{H}_L$ and $\hat{H}_R$. By a judicious choice of $\varphi_{L,0}$ and $\varphi_{R,0}$, the $\phi$'s can also be removed from $\hat{U}_L\hat{\tau}_L\hat{U}_D^\dag$ and $\hat{U}_R\hat{\tau}_R\hat{U}_D^\dag$. All $\phi$'s in $\hat{H}_D$, however, are preserved. It is this type of the gauge transformation that we use in Sec.~\ref{sec:stacking_fault_transmission} to study the effect of the device phases on the transmission function of a single stacking fault.

\section{\label{app:GF_derivation}Derivation of the Green's function}

The evaluation of the Green's function, Eq.~(\ref{eq:chain_eigenfunction_expansion}), in the thermodynamic limit leads to the integral
\begin{equation}
\label{eq:app_g_mn-z-theta-integral}
    g_{mn}(z)=\frac{2}{\pi}\int_0^\pi d\beta\, \frac{\sin(m\beta)\,\sin(n\beta)}{z-\epsilon(\beta)},
\end{equation}
where $m,n$ are integers, $z$ is a complex parameter with the dimension of energy, and $\epsilon(\beta)=\varepsilon_0+2t\cos\beta$. Introducing a dimensionless parameter $\zeta=(z-\varepsilon_0)/2t$, Eq. (\ref{eq:app_g_mn-z-theta-integral}) takes the form
\begin{equation*}
    g_{mn}(\zeta)=\frac{1}{4\pi t}\int_{-\pi}^{\,\pi}\di{\beta}\,\frac{\cos{|m-n|\beta}-\cos{(m+n)\beta}}{\zeta-\cos{\beta}}.
\end{equation*}
We denote the first term on the right-hand side by $g_{|m-n|}$ and the second term by $g_{m+n}$. Below we focus only on $g_{|m-n|}$, as all our subsequent manipulations are equally applicable to $g_{m+n}$. 

Introducing the notation $\xi=e^{i\beta}$ and replacing the domain of integration in $g_{|m-n|}$ with a contour integral over the unit circle $C$, we obtain
\begin{equation*}
    g_{|m-n|}(\zeta)=\frac{i}{2\pi t}\oint_C\di{\xi}\;\frac{\xi^{|m-n|}}{\xi^2-2\zeta\xi+1}.
\end{equation*}
The integrand has simple poles at $\xi_\pm=\zeta\pm\sqrt{\zeta^2-1}$, where the branch cut spans $\zeta\in[-1,1]$ and is so chosen that $\sqrt{\zeta^2-1}$ is positive on the positive imaginary axis. Following the arguments in Ref.~\cite{economou2005}, for $\zeta$ not coinciding with the branch cut, only the pole at $\xi_-$ lies inside $C$. Then, by the residue theorem,
\begin{equation*}
    g_{|m-n|}(\zeta)=\frac{1}{2t}\,\frac{\left[\zeta-\sqrt{\zeta^2-1}\,\right]^{|m-n|}}{\sqrt{\zeta^2-1}}.
\end{equation*}
Combining the contributions from $g_{|m-n|}$ and $g_{m+n}$, we have
\begin{eqnarray*}
    && g_{mn}(\zeta)\\
    && \quad =\frac{1}{2t}\,\frac{\left[\zeta-\sqrt{\zeta^2-1}\,\right]^{|m-n|}-\left[\zeta-\sqrt{\zeta^2-1}\,\right]^{m+n}}{\sqrt{\zeta^2-1}}.
\end{eqnarray*}

We focus on the retarded Green's function within the energy band given by Eq.~(\ref{eq:chain-energy-parametrization}). It is obtained by setting $z=E+i0^+$, where $E\in[\varepsilon_0-2t,\varepsilon_0+2t]$ is real. Choosing the branch cut the same as above and letting $x$ denote the real part of $\zeta=x+i0^+$, we find
\begin{eqnarray*}
    && g_{mn}^{\mathrm{r}}(x)\\
    && \quad =\frac{1}{2it}\,\frac{\left[x-i\sqrt{1-x^2}\,\right]^{|m-n|}-\left[x-i\sqrt{1-x^2}\,\right]^{m+n}}{\sqrt{1-x^2}}.
\end{eqnarray*}
For a chain with periodic boundary conditions, only the translationally-invariant term involving $|m-n|$ appears in the Green's function~\cite{economou2005}. The term involving $(m+n)$ is a consequence of working with open boundary conditions. 

A compact representation of $g_{mn}^{\mathrm{r}}$ is obtained by using the parametrization $x=(E-\varepsilon_0)/2t=\cos{\theta}$, where $0\leq \theta\leq\pi$, which is similar to Eq.~(\ref{eq:chain-energy-parametrization}). Assuming, without loss of generality, that $m\geq n$ (since $g_{mn}^{\mathrm{r}}$ is symmetric in its indices), we find
\begin{equation}
    g_{mn}^{\mathrm{r}}(\theta)=\frac{1}{t}\frac{e^{-im\theta}\sin{\left(n\theta\right)}}{\sin{\theta}},
    \label{eq:I_m,n}
\end{equation}
which is the form of the retarded Green's function given in Eq.~(\ref{eq:chain_retarded_greens_function}).

\section{\label{app:Caroli}Derivation of the Caroli formula}

In this section, we sketch the derivation of the Caroli formula for the transmission function of a non-interacting system, given in Eq.~(\ref{eq:stacking_fault_Caroli_formula}). We follow very closely the approach of  Ref.~\cite{paulsson2006}.

The Schr\"{o}dinger equation of a lead-device-lead system, $\hat H\ket{\psi}=E\ket{\psi}$, may be partitioned into the block-matrix equation
\begin{equation}
    \begin{pmatrix}
        \hat{H}_L&\hat{\tau}_L&0\\[0.3em]
        \hat{\tau}_L^\dag&\hat{H}_D&\hat{\tau}_R^\dag\\[0.3em]
        0&\hat{\tau}_R&\hat{H}_R
    \end{pmatrix}
    \begin{pmatrix}
        \ket{\psi_L}\\[0.3em]\ket{\psi_D}\\[0.3em]\ket{\psi_R}
    \end{pmatrix}
    =E\begin{pmatrix}
        \ket{\psi_L}\\[0.3em]\ket{\psi_D}\\[0.3em]\ket{\psi_R}
    \end{pmatrix},
    \label{eq:app_Caroli_block_Schrodinger}
\end{equation}
where the meanings of each of the submatrices are as in Sec.~\ref{sec:stacking_fault_transmission}, and $\ket{\psi_L}$ is the component of the wavefunction in the left lead, \textit{etc}. We seek to calculate the Green's function of the system, given by
\begin{equation}
    \hat{G}(E)=\bigl(E-\hat H\bigr)^{-1}
    =\begin{pmatrix}
        \hat{G}_{LL}&\hat{G}_{LD}&\hat{G}_{LR}\\[0.3em]
        \hat{G}_{DL}&\hat{G}_{DD}&\hat{G}_{DR}\\[0.3em]
        \hat{G}_{RL}&\hat{G}_{RD}&\hat{G}_{RR}
    \end{pmatrix}.
\label{eq:app_G(E)_definition}
\end{equation}
Here, and in what follows, $\hat{G}(E)$ denotes the Green's function of the \textit{entire} lead-device-lead system, while $\hat{g}_a(E)=(E-\hat{H}_a)^{-1}$ denotes the Green's function of the \textit{isolated} $a^{\mbox{th}}$ lead (when it is disconnected from the device). The index $a\in\{L,R\}$ labels the two leads. In keeping with our usual convention, every Green's function in this section is a \textit{retarded} Green's function, though we shall not designate them as such with a superscript~`$\mathrm{r}$'. 

We obtain, from Eq.~(\ref{eq:app_G(E)_definition}), the following formulas for the relevant blocks of $\hat{G}(E)$:
\begin{subequations}
    \begin{align}
        \hat{G}_{LD}(E)&=\hat{g}_L(E)\,\hat{\tau}_L\,\hat{G}_D(E)
        \label{eq:app_Caroli_GLD},\\
        \hat{G}_{RD}(E)&=\hat{g}_R(E)\,\hat{\tau}_R\,\hat{G}_D(E),
        \label{eq:app_Caroli_GRD}
    \end{align}
\end{subequations}
where $\hat{G}_D(E)$ represents the Green's function of the device \textit{when connected to the leads}, given by
\begin{equation}
    \hat{G}_D(E)=\bigl(E-\hat{H}_D-\hat{\Sigma}_L-\hat{\Sigma}_R\bigr)^{-1}.
    \label{eq:app_Caroli_GD}
\end{equation}
Here, $\hat{\Sigma}_a=\hat{\tau}_a^\dag\,\hat{g}_a\,\hat{\tau}_a$ are the self-energies arising from the device-lead couplings.

Next,  we suppose that rightward-propagating electrons are injected into the system via the left lead. The wavefunction in the left lead will consist of two parts: one wave which is totally reflected from the rightmost surface of the lead (an eigenstate of the isolated left lead), and another corresponding to the ``retarded'' response of the system. These two contributions are denoted $\ket{\psi_{L,n}}$ and $\ket{\psi_L^\mathrm{r}}$, respectively, where $n$ is a collective index of the quantum numbers characterizing the left lead. Similarly, the components of the wavefunction in the subspaces of the device and right lead are denoted $\ket{\psi_D^\mathrm{r}}$ and $\ket{\psi_R^\mathrm{r}}$, respectively. The total wavefunction therefore reads $\ket{\psi}=\left(\ket{\psi_{L,n}},0,0\right)^\top+\ket{\psi^\mathrm{r}}$, where $\ket{\psi^{\mathrm{r}}}=\left(\ket{\psi_{L}^{\mathrm{r}}},\ket{\psi_{D}^{\mathrm{r}}},\ket{\psi_{R}^{\mathrm{r}}}\right)^\top$. Solving the Schr\"{o}dinger equation for $\ket{\psi^\mathrm{r}}$ in terms of $\ket{\psi_{L,n}}$ gives 
\begin{equation}
    \ket{\psi^\mathrm{r}}
    =\hat{G}\begin{pmatrix}
        0\\[0.3em]\hat\tau_L^\dag\ket{\psi_{L,n}}\\[0.3em]0
    \end{pmatrix},
    \label{eq:app_Caroli_retarded_response}
\end{equation} 
from which, using Eqs.~(\ref{eq:app_Caroli_GLD}),~(\ref{eq:app_Caroli_GRD}), and~(\ref{eq:app_Caroli_GD}), one can extract the components of the retarded response in each subspace:
\begin{subequations}
    \begin{align}
        & \ket{\psi_L^\mathrm{r}}
        =\hat{g}_L\hat{\tau}_L\hat{G}_D\hat{\tau}_L^\dag\ket{\psi_{L,n}},
        \label{eq:app_Caroli_psiL_retarded}\\[0.3em]
        & \ket{\psi_D^\mathrm{r}}
        =\hat{G}_D\hat{\tau}_L^\dag\ket{\psi_{L,n}},
        \label{eq:app_Caroli_psiD_retarded}\\[0.3em]
        & \ket{\psi_R^\mathrm{r}}
        =\hat{g}_R\hat{\tau}_R\hat{G}_D\hat{\tau}_L^\dag\ket{\psi_{L,n}}.
        \label{eq:app_Caroli_psiR_retarded}
    \end{align}
\end{subequations}

The final ingredient we need is an expression for the current operator of a discretized Hamiltonian. This can be obtained by demanding that the probability of finding an electron somewhere in the device, $\sum_m|\psi_{D}(m)|^2$, where $m$ runs over the device sites, be conserved in the steady-state. Using the block form of the Hamiltonian in Eq.~(\ref{eq:app_Caroli_block_Schrodinger}), the current transported from the $a^{\mbox{th}}$ lead into the device comes out to be
\begin{equation}
    J_a=-\frac{ie}{\hbar}\left(\braket{\psi_a^\mathrm{r}|\hat{\tau}_a|\psi_D^\mathrm{r}}-\braket{\psi_D^\mathrm{r}|\hat{\tau}_a^\dag|\psi_a^\mathrm{r}}\right).
    \label{eq:app_Caroli_current}
\end{equation}
Finally, to obtain the Caroli formula, we use Eqs.~(\ref{eq:app_Caroli_psiL_retarded}) through~(\ref{eq:app_Caroli_psiR_retarded}) to express all retarded responses in Eq.~(\ref{eq:app_Caroli_current}) in terms of $\ket{\psi_{L,n}}$. This gives the contribution to the total current transmitted through the device and into the right lead due to the $n^{\mbox{th}}$ mode in the left lead. We then sum over all $n$ and integrate over all incident energies $E$ which are such that $E=E_n$, where $E_n$ is the energy level of the $n^{\mbox{th}}$ mode; the result is
\begin{equation}
    J_{L\rightarrow R}
    =\frac{2e}{h}\int\di{E}\,f(E,\mu_L)\,\Tr{\hat{G}_D^\dag\hat{\Gamma}_R\hat{G}_D\hat{\Gamma}_L},
\end{equation}
where $\hat{\Gamma}_a=-2\im{\hat{\Sigma}_a}$ are the level-width functions and $f(E,\mu_L)$ is the Fermi-Dirac distribution characterizing the left lead. The quantity in the square brackets is precisely the Caroli formula, Eq.~(\ref{eq:stacking_fault_Caroli_formula}).

\section{\label{app:recursion}Derivation of Eq.~(\ref{eq:random_chain_unperturbed_device_greens_function}).}

Let $\hat{T}$ be the $M\times M$ tridiagonal matrix\begin{equation}
    \hat{T}=\begin{pmatrix}
        a_{1}&b_{1}\\
        c_{1}&a_{2}&b_{2}\\
        &c_{2}&\ddots &\ddots \\
        &&\ddots &\ddots &b_{n-1}\\
        &&&c_{n-1}&a_{n}
    \end{pmatrix}.
    \label{eq:recursion_tridiagonal_matrix}
\end{equation}
According to Ref.~\cite{usmani_1994} the inverse matrix has the following elements:
\begin{equation}
    T^{-1}_{mn}=\frac{1}{\xi_M}
    \begin{cases}
        (-1)^{m+n}\left[\prod_{r=m}^{n-1}b_r\right]\xi_{m-1}\eta_{n+1},&m<n,\\[0.3em]
        \xi_{m-1}\eta_{n+1},&m=n,\\[0.3em]
        (-1)^{m+n}\left[\prod_{r=n}^{m-1}c_r\right]\xi_{n-1}\eta_{m+1},&m>n.
    \end{cases}
    \label{eq:recursion_inverse_matrix_elements}
\end{equation}
The auxiliary variables $\xi_m,\eta_m$ must be chosen to satisfy the recurrence relations
\begin{subequations}
\begin{alignat}{2}
    \xi_m&=a_{m}\xi_{m-1}-b_{m-1}c_{m-1}\xi_{m-2},\quad m\in[2,M],
    \label{eq:recursion_xi_recurrence}\\[0.3em]
    \eta_m&=a_{m}\eta_{m+1}-b_{m}c_{m}\xi_{m+2},\quad m\in[M-1,1],
    \label{eq:recursion_eta_recurrence}
\end{alignat}
\end{subequations}
subject to the boundary conditions
\begin{alignat}{2}
    \xi_0=1,\quad
    \xi_1=a_1,\quad
    \eta_{M+1}=1,\quad
    \eta_M=a_M.
    \label{eq:recursion_BCs}
\end{alignat}

The matrix we need to invert is given by Eq.~(\ref{eq:random_chain_device_greens_function_definition}); namely, all off-diagonal elements $b_m$, $c_m$ are constant and equal to $t$. All diagonal elements $a_m$ are equal to $2t\cos\theta$, with the exceptions of the $(1,1)$ and $(M,M)$ elements, which are equal to $te^{i\theta}$. It may be verified by induction, with this set of $a_m$, $b_m$, and $c_m$, that the solutions to Eqs.~(\ref{eq:recursion_xi_recurrence}) and~(\ref{eq:recursion_eta_recurrence}) which are constrained according to Eq.~(\ref{eq:recursion_BCs}) have the form
\begin{subequations}
\begin{align}
    \xi_m&=\left\{\begin{array}{ll}
         t^me^{im\theta},&m\in[0,M-1],\\
        2it^Me^{i(M-1)\theta}\sin\theta,&m=M
    \end{array}\right.,
    \label{eq:recursion_xi_solution}\\[0.3em]
    \eta_m&=\left\{\begin{array}{ll}
        t^{M-m+1}e^{i(M-m+1)\theta},&m\in[M+1,2],\\
        2it^Me^{iM\theta}\sin\theta,&m=1.
    \end{array}\right.
    \label{eq:recursion_eta_solution}
\end{align}
\end{subequations}
Substitution of Eqs.~(\ref{eq:recursion_xi_solution}) and (\ref{eq:recursion_eta_solution}) into Eq. (\ref{eq:recursion_inverse_matrix_elements}) produces, after some algebra, Eq. (\ref{eq:random_chain_unperturbed_device_greens_function}).

\section{\label{app:stacking_fault}A different stacking fault}

\begin{figure}[t]
    \centering
    \begin{tikzpicture}
    \def\a{1.1}
    \foreach \i in {1,...,4} {\draw (\a*\i,0) circle (0.2) node {$\i$};}
    \node at ({-1*\a},0) {$\cdots$};
    \node at ({6*\a},0) {$\cdots$};
    \foreach \i in {0,5} {\fill (\a*\i,0) circle (0.2) node [color=white] {$\i$};}
    \draw [gray,->,>=stealth,very thin] ({0*\a+0.1},0.3) to [out=45,in=135] node [midway, above] {$te^{i\phi_0}$} ({1*\a-0.1},0.3);
    \draw [gray,->,>=stealth,very thin] ({1*\a+0.1},0.3) to [out=45,in=135] node [midway, above] {$te^{i\phi_1}$} ({2*\a-0.1},0.3);
    \draw [gray,->,>=stealth,very thin] ({2*\a+0.1},0.3) to [out=45,in=135] node [midway, above] {$te^{i\phi_2}$} ({3*\a-0.1},0.3);
    \draw [gray,->,>=stealth,very thin] ({3*\a+0.1},0.3) to [out=45,in=135] node [midway, above] {$te^{i\phi_3}$} ({4*\a-0.1},0.3);
    \draw [gray,->,>=stealth,very thin] ({4*\a+0.1},0.3) to [out=45,in=135] node [midway, above] {$te^{i\phi_4}$} ({5*\a-0.1},0.3);
    \draw [gray,->,>=stealth,very thin] ({2*\a},-0.3) --++ (0,-0.8) --++ ({2*\a},0) node [fill=white,midway] {$t'e^{i\alpha_2}$} --++ (0,0.8);
    \draw [gray,->,>=stealth,very thin] ({1*\a},-0.3) --++ (0,-0.4) --++ ({2*\a},0) node [fill=white,midway] {$t'e^{i\alpha_1}$} --++ (0,0.4);
\end{tikzpicture}
    \caption{Lead-device-lead arrangement for the configuration described in Eq.~(E1) (shown here in the effective~1D chain picture). The device and lead sites are represented by open and closed circles, respectively. First-neighbour hoppings with phases $\phi_j$ exist between all sites in the system, while second-neighbour hoppings occur only within the device.}
    \label{fig:app_stacking_fault}
\end{figure}
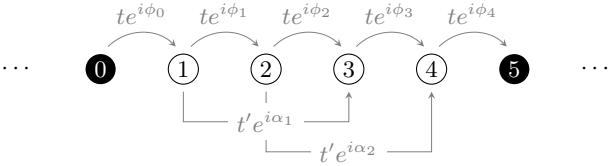

In Sec.~\ref{sec:stacking_fault_transmission}, we considered a stacking fault described by the left and right leads meeting at a ``domain wall'' (see Fig.~\ref{fig:stacking_fault}). This is not the only type of stacking fault that we might encounter. Here, we calculate the transmission function for a different stacking fault. We consider a single~`$-$' defect in the H\"{a}gg code, given by
\begin{subequations}
    \begin{align}
        & \mbox{Barlow:}\quad
        \cdots CAB\lvert CACA\rvert BCA\cdots
        \label{eq:app_stacking_fault_barlow_sequence}\\
        & \mbox{H\"{a}gg:}\quad
        \left(\cdots++\right)+\left[+-+\right]+\left(++\cdots\right).
        \label{eq:app_stacking_fault_hagg_code}
    \end{align}
\end{subequations}
The two sign changes in Eq.~(\ref{eq:app_stacking_fault_hagg_code}), $(+-)$ and $(-+)$, now give rise to two~3nn hoppings between the first and third, and the  second and fourth layers in the device (see Fig.~\ref{fig:app_stacking_fault} for an illustration). The device Hamiltonian reads
\begin{equation}
    \hat{H}_D=\begin{pmatrix}
        \varepsilon_0&-te^{i\phi_1}&-t'e^{i\alpha_1}&0\\[0.3em]
        -te^{-i\phi_1}&\varepsilon_0&-te^{i\phi_2}&-t'e^{i\alpha_2}\\[0.3em]
        -t'e^{-i\alpha_1}&-te^{-i\phi_2}&\varepsilon_0&-te^{i\phi_3}\\[0.3em]
        0&-t'e^{-i\alpha_2}&-te^{-i\phi_3}&\varepsilon_0
    \end{pmatrix}.
    \label{eq:app_stacking_fault_device_hamiltonian}
\end{equation}
As in Sec.~\ref{sec:stacking_fault_transmission}, we gauge away all phases from the leads and allow the $\phi$'s and $\alpha$'s in Eq.~(\ref{eq:app_stacking_fault_device_hamiltonian}) to be arbitrary real numbers.

Generalizing Eq.~(\ref{eq:stacking_fault_Caroli_formula_simplified}) to a four-layer device, and using the self-energies and level-width functions from Sec.~\ref{sec:stacking_fault_transmission} [see Eqs.~(\ref{eq:stacking_fault_sigma_L}) through~(\ref{eq:stacking_fault_gamma_R})], we find that the transmission function is
\begin{widetext}
\begin{equation}
    T_{1\mathrm{D}}(E)=\left|\frac{2\sin{\theta}\left(\Delta^2e^{i\Phi_2}-2\Delta\cos{\theta}\left[1+e^{i\left(\Phi_2-\Phi_1\right)}\right]+e^{-i\Phi_1}\right)}{\Delta^4-2\Delta^2\left[2e^{i\theta}\cos{\theta}+\cos{\left(\Phi_1-\Phi_2\right)}\right]+2\Delta e^{i\theta}\left(\cos{\Phi_1}+\cos{\Phi_2}\right)+2ie^{3i\theta}\sin{\theta}}\right|^{2},
    \label{eq:app_stacking_fault_transmission_function}
\end{equation}
\end{widetext}
where $\Delta=t'/t$, angle $\theta$ is given implicitly in terms of $E$ through Eq.~(\ref{eq:chain-energy-parametrization}). We have introduced the new phases
\begin{equation}
    \Phi_1=\phi_1+\phi_2-\alpha_1,\quad
    \Phi_2=\phi_2+\phi_3-\alpha_2,
\end{equation}
which correspond to the net ``magnetic fluxes'' through each of the two closed loops in Fig.~\ref{fig:app_stacking_fault}. Again, for the physically relevant case of Eq.~(\ref{eq:close_packed_transcriptions}), the phases satisfy $\phi_1=-\phi_2=\phi_3$, while $\alpha_1=\alpha_2=0$, so that $\Phi_1=\Phi_2=0$, and Eq.~(\ref{eq:app_stacking_fault_transmission_function}) somewhat simplifies. The key point is that, without $t'$, the transmission function is unaffected by phase disorder (and equal to one).

For $\Delta\ll1$ and far from either band edge, Eq.~(\ref{eq:app_stacking_fault_transmission_function}) becomes
\begin{equation}
    T(E)\simeq 1-\frac{4\Delta^2}{\sin^2{\theta}},
    \label{eq:app_stacking_fault_transmission_function_asymptotic}
\end{equation}
which has a similar form to the expression (\ref{eq:stacking_fault_transmission_function_asymptotic}) for the transmission function of the ``domain wall''-type stacking fault.

\section{\label{app:BZ_integration}Brillouin zone integration}

The overall transmission function of the sample is obtained by integrating $T(\kp,E)$ over all $\kp$-channels:
\begin{equation}
    T(E)
    =\sum_{\kp}T(\kp,E)
    =A\int_{\Omega(E)}\frac{\di^2\kp}{(2\pi)^2}\:T(\kp,E).
\label{eq:overall-T-dimensional-k}
\end{equation}
Here, $A$ is the cross-sectional area of the sample, and $\kp=(k_x,k_y)$ is the in-plane momentum expressed in the Cartesian basis. We let $a$ denote the interatomic spacing. In this Appendix, we do \textit{not} use units in which $a=1$. Thus, $k_x$ and $k_y$ have the dimension of inverse length. We have $A=\mathcal{N}A_{\mathrm{cell}}$, where $\mathcal{N}$ is the number of unit cells per layer, and $A_{\mathrm{cell}}=\sqrt{3}a^2/2$ is the area of a single unit cell.

It is convenient to re-express $\kp$ as
\begin{equation}
    \kp=\frac{\kappa_a}{2\pi}\bm{K}_a+\frac{\kappa_b}{2\pi}\bm{K}_b
\end{equation}
where the $\bm{K}$'s are the reciprocal lattice vectors corresponding to the primitive lattice vectors in Eq.~(\ref{eq:lattice_vectors}), and the $\kappa$'s are dimensionless numbers whose range is restricted such that $\kp$ runs over the first Brillouin zone. It is easy to show that the Jacobian associated with the change of variables from $\{k_x,k_y\}$ to $\{\kappa_a,\kappa_b\}$ is equal to $2/\sqrt{3}a^2=A_{\mathrm{cell}}^{-1}$.

Putting everything together, the transmission function, Eq. (\ref{eq:overall-T-dimensional-k}), can be written as 
\begin{equation}
    T(E)
    =\mathcal{N}\int_{\Omega(E)}\frac{\di\kappa_a\di\kappa_b}{(2\pi)^2}\:T(\kappa_a,\kappa_b,E)
    \equiv \mathcal{N}T_{\mathrm{cell}}(E).
\end{equation}
From the Landauer formula, Eq. (\ref{eq:momentum_average_resistance_def}), we obtain:
\begin{equation}
    R_{\mathrm{cell}}=R\mathcal{N}=\frac{h}{2e^2}\,\frac{1}{T_{\mathrm{cell}}(E_F)}.
\end{equation}
Here, $R$ is the resistance of an entire layer transverse to the stacking direction, while $R_{\mathrm{cell}}$ can be interpreted as the resistance of a single unit cell.

Finally, to return to more experimentally-accessible units, we replace $\mathcal{N}$ by its definition in terms of $A$ and $A_{\mathrm{cell}}$, which yields
\begin{equation}
    r_S=RA=\frac{h}{2e^2}\,\frac{A_{\mathrm{cell}}}{T_{\mathrm{cell}}(E_F)}.
    \label{eq:app_BZ_specific_resistance}
\end{equation}
The left-hand side now has the dimension of resistance multiplied by area, and is interpreted as the ``specific'' resistance of a stacking fault (cf. Ref.~\cite{Koren2014}).

\bibliography{main}

\end{document}